\documentclass[fleqn,usenatbib]{mnras}

\usepackage{newtxtext,newtxmath}
\usepackage[T1]{fontenc}
\usepackage{lscape}
\usepackage{longtable}
\usepackage{booktabs}
\usepackage{multirow}
\usepackage{graphicx}	% Including figure files
\usepackage{amsmath}	% Advanced maths commands
\usepackage{amssymb}	% Extra maths symbols
\usepackage{enumitem}
\usepackage{xcolor}
\colorlet{RED}{red}
\setlist[enumerate]{
  labelsep=8pt,
  labelindent=0\parindent,
  itemindent=0pt,
  leftmargin=*,
}

\title[SNAD: SuperNova Anomaly Detection]{Anomaly Detection in the Open Supernova Catalog}

\author[M. V. Pruzhinskaya et al.]{M.~V.~Pruzhinskaya,$^{1}$\thanks{E-mail: pruzhinskaya@gmail.com}
K.~L.~Malanchev,$^{1,2}$\thanks{E-mail: malanchev@physics.msu.ru}
M.~V.~Kornilov,$^{1,2}$
E.~E.~O.~Ishida,$^{3}$ 
F.~Mondon,$^{3}$
\newauthor
A.~A.~Volnova$^{4}$
and V.~S.~Korolev$^{5,6}$
\\
$^{1}$Lomonosov Moscow State University, Sternberg Astronomical Institute, Universitetsky pr.~13, Moscow, 119234, Russia\\
$^{2}$National Research University Higher School of Economics, 21/4 Staraya Basmannaya Ulitsa, Moscow, 105066, Russia\\
$^{3}$Universit\'e Clermont Auvergne, CNRS/IN2P3, LPC, F-63000 Clermont-Ferrand, France\\
$^{4}$Space Research Institute of the Russian Academy of Sciences (IKI), 84/32 Profsoyuznaya Street, Moscow, 117997, Russia\\
$^{5}$Central Aerohydrodynamic Institute, 1 Zhukovsky st, Zhukovsky, Moscow Region, 140180, Russia\\
$^{6}$Moscow Institute of Physics and Technology, 9 Institutskiy per., Dolgoprudny, Moscow Region, 141701, Russia
}

\date{Accepted XXX. Received YYY; in original form ZZZ}

\pubyear{2019}

\begin{document}
\graphicspath{{figures/}}
\label{firstpage}
\pagerange{\pageref{firstpage}--\pageref{lastpage}}

\maketitle

% Abstract of the paper
\begin{abstract}
In the upcoming decade large astronomical surveys will discover millions of transients raising unprecedented data challenges in the process. Only the use of the machine learning algorithms can process such large data volumes. Most of the discovered transients will belong to the known classes of astronomical objects. However, it is expected that some transients will be rare or completely new events of unknown physical nature. The task of finding them can be framed as an anomaly detection problem. In this work, we perform for the first time an automated anomaly detection analysis in the photometric data of the Open Supernova Catalog (OSC), which serves as a proof of concept for the applicability of these methods to future large scale surveys. The analysis consists of the following steps: 1) data selection from the OSC and approximation of the pre-processed data with Gaussian processes, 2) dimensionality reduction, 3)  searching for outliers with the use of the isolation forest algorithm, 4) expert analysis of the identified outliers.  The pipeline returned 81 candidate anomalies, 27 (33\%) of which were confirmed to be from astrophysically peculiar objects. Found anomalies correspond to a selected sample of 1.4\% of the initial automatically identified data sample of $\sim$2000 objects. Among the identified outliers we recognised superluminous supernovae, non-classical Type~Ia supernovae, unusual Type~II supernovae, one active galactic nucleus and one binary microlensing event. We also found that 16 anomalies classified as supernovae in the literature are likely to be quasars or stars. Our proposed pipeline represents an effective strategy to guarantee we shall not overlook exciting new science hidden in the data we fought so hard to acquire. All code and products of this investigation are made publicly available\footnotemark. 
\end{abstract}

%%%%%%%%%%%%%%%%%%%%%%%%%%%%%%%%%%%%%%%%%%%%%%%%%%

%%%%%%%%%%%%%%%%% BODY OF PAPER %%%%%%%%%%%%%%%%%%
\footnotetext{\href{http://snad.space/osc/}{http://snad.space/osc/}}

\clearpage
\section{Introduction}
\label{sec:introduction}

Supernovae (SNe) hold vital pieces of the large cosmic puzzle astronomy and cosmology aim to solve. They are responsible for the chemical enrichment of interstellar medium~\citep{2013ARA&A..51..457N}; the production of high energy cosmic rays~\citep{2017hsn..book.1711M}, and they trigger star formation via the density waves induced by their energetic explosions~\citep{Nagakura2009,2013ApJ...762...50C}. Moreover, the study of different types of SNe allows us to probe the composition and distance scale of the Universe~\citep{1974ApJ...193...27K,2002ApJ...566L..63H,Riess98,Perlmutter99} --- imposing strong constraints on the standard cosmological model~\citep{Betoule2014,2018ApJ...859..101S}.

Given the potential impact of SN research on different areas of astronomy, the scientific community has allocated a large fraction of its efforts in the generation of large supernova surveys --- a few recent examples include the Carnegie Supernova Project\footnote{\href{https://csp.obs.carnegiescience.edu/}{https://csp.obs.carnegiescience.edu/}} (CSP;~\citealt{2006PASP..118....2H}), the Panoramic Survey Telescope and Rapid Response System\footnote{\href{https://panstarrs.stsci.edu/}{https://panstarrs.stsci.edu/}} (Pan-STARRS;~\citealt{2010SPIE.7733E..0EK,2016arXiv161205560C}), the Dark Energy Survey\footnote{\href{https://www.darkenergysurvey.org/}{https://www.darkenergysurvey.org/}}~\citep[DES;][]{2016MNRAS.460.1270D} and the Zwicky Transient Facility\footnote{\href{https://www.ztf.caltech.edu/}{https://www.ztf.caltech.edu/}} (ZTF;~\citealt{2019PASP..131a8002B}). Another generation of even larger counterparts, like the Large Synoptic Survey Telescope\footnote{\href{https://www.lsst.org/}{https://www.lsst.org/}} (LSST;~\citealt{2009arXiv0912.0201L}), will soon join this list, making available a combined data set of  unprecedented volume and complexity. 

In this new data paradigm, the use of machine learning (ML) methods is unavoidable \citep{2010IJMPD..19.1049B}. Astronomers have already benefited from developments in machine learning, in particular for exoplanet search~\citep{2015ApJ...806....6M,2015ApJ...812...46T,2018MNRAS.474..478P}, but the synergy is far from that achieved by other endeavours in genetics~\citep{CHEN2012323,Libbrecht15,10.1093/nar/gkw226}, ecology~\citep{Criscia12} or medicine~\citep{VENKATRAGHAVAN2019518,DUBOST2019534}. Moreover, given the relatively recent advent of large data sets, most of the ML efforts in astronomy are concentrated in classification~\citep[e.g.,][]{2010PASP..122.1415K,2013MNRAS.430..509I,2016ApJS..225...31L,2016ApJ...821...86H,2019MNRAS.483....2I,2018arXiv181108446S} and regression~\citep[e.g.,][]{2010A&A...523A..31H,2015MNRAS.452.3100C,8285192,2017MNRAS.468.4323B} tasks. Machine learning is also actively applied for the real-bogus classification that allows to automatically disentangle real transients from the artefacts on the images produced by major time-domain surveys~\citep{2012PASP..124.1175B,2015MNRAS.449..451W,2015AJ....150...82G,2015MNRAS.454.2026D}. A large variety of ML  methods were applied to supervised photometric SN classification problem   \citep{2012MNRAS.419.1121R,2015ApJ...800...36S,2016ApJS..225...31L,2016JCAP...12..008M,2017ApJ...837L..28C,2018MNRAS.473.3969R,2019arXiv190100461B,2019arXiv190101298P,2019arXiv190106384M} and unsupervised characterisation from spectroscopic observation \citep[e.g., ][]{2016ApJ...828..111R,2016MNRAS.461.2044S,2019arXiv190302557M}.

Astronomical anomaly detection has not been yet fully implemented in the enormous amount of data that has been gathered. Barring a few exceptions, most of the previous studies can be divided into only two different trends: clustering~\citep[e.g., ][]{Rebbapragada2009} and subspace analysis~\citep[e.g., ][]{doi:10.1002/sam.11167} methods. More recently, random forest algorithms have been extensively used by themselves~\citep{2017MNRAS.465.4530B} or in hybrid statistical analysis~\citep{2014ApJ...793...23N}. Although all of this has been done to periodic variables there is not much done for transients and even less for supernovae. 

The lack of spectroscopic support causes the large supernova databases to collect SN candidates basing on the secondary indicators (proximity to the galaxy, arise/decline rate on a light curve (LC), absolute magnitude). This leads to the appearance of incorrectly classified objects. Anomaly detection can help us to purify the supernova databases from the non-supernova contamination. It is also expected that during such analysis the unknown variable objects or SNe with unusual properties can be detected. As an example of unique objects one can refer to SN2006jc --- SN with very strong but relatively narrow He~I lines in early spectra ($\sim$30 similar objects are known,~\citealt{2016MNRAS.456..853P}), SN2005bf --- supernova attributed to SN~Ib but with two broad maxima on LCs~\citep{2006ApJ...641.1039F}, SN2010mb --- unusual SN~Ic with very low decline rate after the maximum brightness that is not consistent with radioactive decay of $^{56}$Ni~\citep{2014ApJ...785...37B}, ASASSN-15lh --- for some time it was considered as the most luminous supernova ever observed --- two times brighter than superluminous supernovae (SLSN), later the origin of this object was challenged and now it is considered as a tidal disruption of a main-sequence star by a black hole~\citep{2016Sci...351..257D,2016NatAs...1E...2L}. As such sources are typically rare, the task of finding them can be framed as an anomaly detection problem. 

In this paper we turn to the automatic search for anomalies in the real photometric data using the Open Supernova Catalog\footnote{\href{https://sne.space/}{https://sne.space/}}~(OSC,~\citealt{2017ApJ...835...64G}). The OSC has never been used for the task of the anomaly detection with the ML algorithms until this work, however, it was used for the classification problem~\citep{2018ApJS..236....9N,2019arXiv190302557M}. The anomalies we are looking for are any artefacts in the data, cases of misclassification (active galactic nuclei (AGN), novae, binary microlensing events), rare classes of objects (SLSN, kilonovae, SNe associated with gamma-ray bursts), and objects of unknown nature. We use the isolation forest as an outlier detection algorithm that identifies outliers instead of normal observations~\citep{Liu:2012:IAD:2133360.2133363}. This technique is based on the fact that outliers are data points that are few and different. Similarly to random forest it is built on an ensemble of binary (isolation) trees. The final goal of the presented work is to develop some approach that allows to detect anomalies in huge amount of data produced by time-domain surveys such as LSST. Due to the initial absence of any labelled data in transient databases, the  algorithm  follows the paradigm of unsupervised learning. For this reason we pretend that we do not have any labels in the OSC and we use only the multicolour photometry. Moreover, the spectral classification provided by the OSC is collected from different sources, including the preliminary classification from the astronomical telegrams\footnote{\href{http://www.astronomerstelegram.org/}{http://www.astronomerstelegram.org/}} where it can be based on one spectrum only, that is simply fitted by \textsc{SNID}~\citep{SNID} to the closest supernova template. Such rough classification can not give an information about peculiar behaviour of the source, usually more detailed study is needed. On the contrary, it is not necessary that all outliers found by machine are real anomalies. That is why we also subject the outliers to the careful astrophysical analysis using the publicly available information. 

The rest of the paper is organised as follows.
In Section~\ref{sec:data} we describe the data used for the analysis. Section~\ref{sec:pre-proc} is devoted to work related to the data pre-processing, including light curve approximation by Gaussian processes (GP). The outlier detection algorithm is presented in Section~\ref{sec:iso_forest}. Section~\ref{sec:results} shows the results and contains the analysis of found outliers. We conclude the paper in Section~\ref{sec:conclusions}. The outliers are listed in Appendix~\ref{anomaly_list}.

\begin{figure*}
\begin{center}
\includegraphics[trim={0 15cm 0 0},clip,scale=0.65]{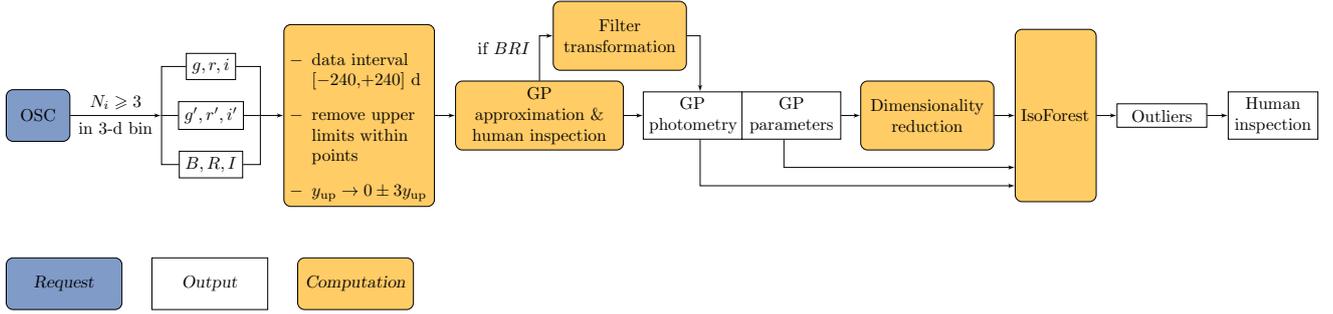}
\caption{Workflow for the analysis. $N_i$ denotes the number of observations in $i$'th band. GP photometry includes 364 features: $121 \times 3$ normalized fluxes and the LC flux maximum; GP parameters are 9 fitted parameters of the Gaussian process kernel and the log-likelihood of the fit.}
\label{workflow}
\end{center}
\end{figure*}

\section{The Open Supernova Catalog}
\label{sec:data}

The data are drawn from the Open Supernova Catalog~\citep{2017ApJ...835...64G}. The catalog is constructed by combining many publicly available data sources such as the Asiago Supernova Catalog~\citep{1999A&AS..139..531B}, the Gaia Photometric Science Alerts~\citep{2012gfss.conf...21W,2014htu..conf...43C},  the Nearby Supernova Factory~\citep{Aldering02}, Pan-STARRS~\citep{2010SPIE.7733E..0EK,2016arXiv161205560C}, the SDSS Supernova Survey~\citep{Sako2018},  the Sternberg Astronomical Institute Supernova Light Curve Catalogue~\citep{2005yCat.2256....0T}, the Supernova Legacy Survey ~\citep[SNLS, ][]{2005ASPC..339...60P,2006A&A...447...31A}, the MASTER Global Robotic Net~\citep{2010AdAst2010E..30L}, the All-Sky Automated Survey for Supernovae~(ASAS-SN, \citealt{2019MNRAS.484.1899H}), and the intermediate Palomar Transient Factory (iPTF, \citealt{2009PASP..121.1395L,2016PASP..128k4502C}) among others, as well as from individual publications.  It represents an open repository for supernova metadata, light curves, and spectra in an easily downloadable format. This catalog also includes some contamination from non-SN objects.

Given the large number of objects and their diverse characteristics, this catalog is ideal for our goal of automatically identifying anomalies. It incorporates data for more than $5\times10^4$ SNe candidates among which $\sim$1.2$\times10^4$ objects have $>$10 photometric observations and $\sim$5$\times10^3$ have spectra. For comparison, SDSS supernova catalog contains only 4607 SNe candidates: 889 with measured spectra~\citep{Sako2018}.

The catalog stores the data in different photometric passbands. To have a more homogeneous sample, we chose only those objects that have LCs in $BRI$~\citep{1990PASP..102.1181B}, $g'r'i'$ or $gri$ filters. The primed system $u'g'r'i'z'$ is defined  in  the  natural  system  of  the  USNO 1-m telescope. The SDSS magnitudes $ugriz$, however, are defined in the natural system of the SDSS 2.5-m telescope. These two systems are very similar and the coefficients of the transformation equations are quite small~\citep{1996AJ....111.1748F,2006AN....327..821T,2007ASPC..364...91S}. We assume that $g'r'i'$ filters are close enough to $gri$ and transform $BRI$ to $gri$ (see Sect.~\ref{transformation}). We require a minimum of three photometric points in each filter with a 3-day binning (Fig.~\ref{workflow}). Our experiments show that this threshold is enough to provide a good reconstruction of the light curve --- specially in cases where photometric points are not homogeneously distributed among filters. This is natural consequence of the light curve approximation procedure we adopted (Section~\ref{GP}) which takes into account the correlation between photometric bands to guide the reconstruction in sparsely populated filters. After this first cut, our sample consists of 3197 objects (2026 objects in $g'r'i'$, 767 objects in $gri$, and 404 objects in $BRI$).

We downloaded the data from the GitHub page\footnote{\href{https://github.com/astrocatalogs/}{https://github.com/astrocatalogs/}} of the Astrocats project on June, 2018. The complete data set of 45162 objects is located at \href{http://snad.space/osc/sne.tar.lzma}{http://snad.space/osc/sne.tar.lzma}.

\section{Pre-processing}
\label{sec:pre-proc}

In this section we describe how to get features for ML from the OSC light curves. The pre-processing procedure includes several steps that are described in detail in the subsections below and illustrated by Fig.~\ref{workflow}.  First, we prepared the photometric data extracted from the OSC; we transformed the magnitudes to the flux units, converted the upper limits, and implemented 1-day time-binning. Then, we used the Gaussian processes to approximate the photometric observations in each filter. The objects with bad light curve approximations were removed from the further analysis.  After that, we transformed the remaining light curves in $BRI$ filters to $gri$. To have a homogeneous input data, for each object we extracted its photometry in the range $[-20,+100]$ days relative to the maximum flux. We also kept the kernel parameters of the Gaussian processes. All of this together was subjected to the dimensionality reduction procedure using t-SNE method~\citep{maaten_hinton_2008}.

\subsection{Filter transformation}
\label{transformation}
In order to ensure maximum exploitation of the data at hand, we convert the Bessel's $BRI$ into $gri$ filters using the Lupton's (2005) transformation equations\footnote{\href{http://www.sdss3.org/dr8/algorithms/sdssUBVRITransform.php}{http://www.sdss3.org/dr8/algorithms/sdssUBVRITransform.php}}. These equations are derived by matching SDSS DR4 photometry to Peter Stetson's published photometry for stars\footnote{\href{http://www.cadc-ccda.hia-iha.nrc-cnrc.gc.ca/en/community/STETSON/index.html}{http://www.cadc-ccda.hia-iha.nrc-cnrc.gc.ca/en/community/STETSON/index.html}}:

\begin{equation}
 \left\{
      \begin{aligned}
    B &= u - 0.8116 \ (u - g) + 0.1313 \\
    B &= g + 0.3130 \ (g - r) + 0.2271 \\
    V &= g - 0.2906 \ (u - g) + 0.0885\\
    V &= g - 0.5784 \ (g - r) - 0.0038\\
    R &= r - 0.1837 \ (g - r) - 0.0971\\
    R &= r - 0.2936 \ (r - i) - 0.1439\\
    I &= r - 1.2444 \ (r - i) - 0.3820\\
    I &= i - 0.3780 \ (i - z) - 0.3974 
      \end{aligned}
    \right.
    \label{lupton}
\end{equation}

\begin{figure*}
\begin{center}
\includegraphics[scale=0.58]{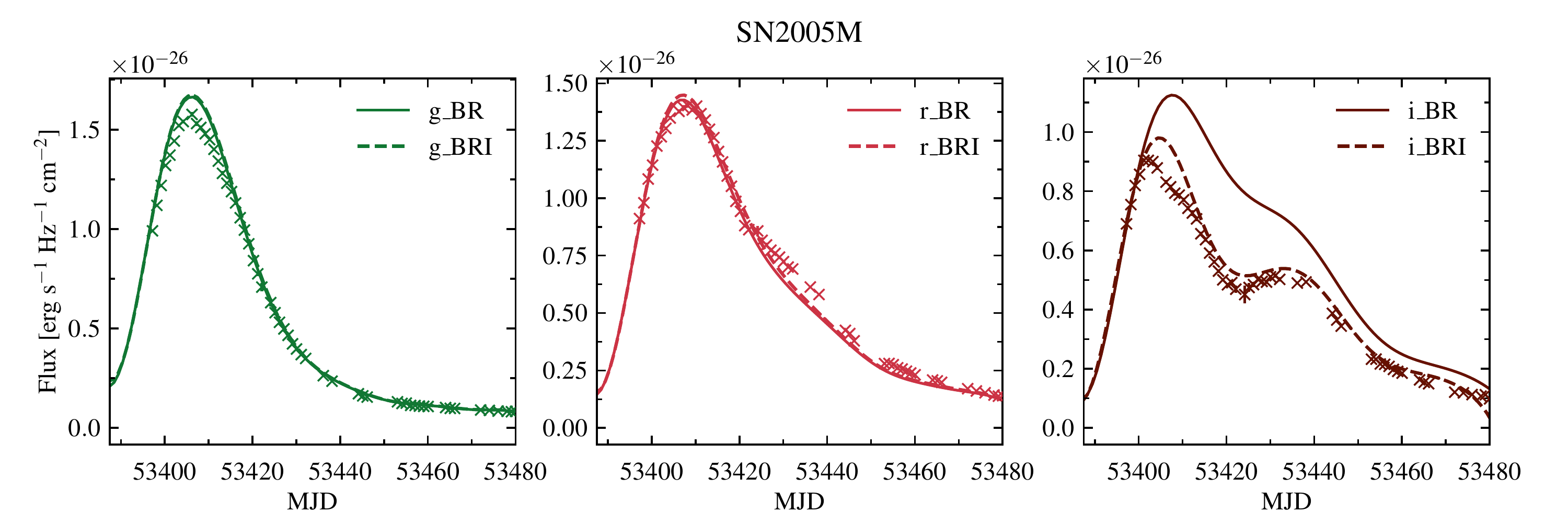}
\caption{Light curves of SN2005M. Crosses are the observations in $gri$ filters~\citep{2010ApJS..190..418G,2010AJ....139..519C,2012MNRAS.425.1789S}. Solid and dashed lines are the approximated and transformed light curves from the Bessel's $BR$ and $BRI$ to $gri$ filters, respectively.}
\label{filt_tr}
\end{center}
\end{figure*}

As we can see, there are several possibilities to obtain $gri$ light curves from the Bessel's ones. Obviously, the accuracy of the transformation increases with the number of available filters. However, in the Open Supernova Catalog objects having photometry only in two filters are more numerous than those having photometry in three or four filters. Therefore, the less filters we use, the larger sample of SNe candidates we obtain. First, we tried to use only two Bessel's filters. Prior to applying the filter transformation, we approximated the LCs with Gaussian processes~(see Sect.~\ref{GP}). To evaluate the quality of transformation, with two filters only, we chose a few objects with LCs available in both, Sloan and Bessel's filters, and compared the transformed $gri$ with the original ones. As can be seen from the Fig.~\ref{filt_tr}, the results of comparison are unsatisfactory for $i$ filter. This indicated that at least one more filter had to be added in the analysis. The same test showed that three filters ($BRI$) are enough to adequately reproduce $gri$ light curves (Fig.~\ref{filt_tr}). Since with 3 filters the equations become over-determined, we used the least-square method to solve~Eq. \ref{lupton}.

Despite the fact that the transformation between the filters depends on the spectrum of an object, and Lupton's equations are derived for stars, not for supernovae, the Fig.~\ref{filt_tr} shows quite good agreement between transformed and original light curves.

\subsection{Light curve approximation}
\label{GP}

Traditionally, ML algorithms require a homogeneous input data matrix which, unfortunately, is not the case with supernovae. A commonly used technique to transform unevenly distributed data into an uniform grid is to approximate them with Gaussian processes~\citep{RasmussenGP}. Usually, each light curve is approximated by GP independently. However, in this study we use a  \textsc{Multivariate Gaussian Process\footnote{\href{https://github.com/matwey/gp-multistate-kernel}{https://github.com/matwey/gp-multistate-kernel}}} approximation. For each object it takes into account the correlation between light curves in different bands, approximating the data by GP in all filters in a one global fit (for details see Kornilov et al. 2019, in prep.). With this technique we can reconstruct the missing parts of LC from its behaviour in other filters. For example, in Fig.~\ref{SN2013cv} maximum in $g$ filter is reproduced from the $r,i$ light curves. This correlation does not rely on any physical assumptions about LC shape. As an approximation range we chose $[-20, +100]$ days. We also extrapolated the GP approximation to fill this range if needed. Once the GP approximation becomes negative, it is zeroed till infinity.

Gaussian process is based on the so-called kernel, a function describing the  covariance between two observations. The kernel used in our implementation of  \textsc{Multivariate Gaussian Process} is composed of three  radial-basis functions $k_i(t_1, t_2) = \exp{\left(-\frac{(t_2-t_1)^2}{2\,l_i^2}\right)}$, where $i$ denotes the photometric band, and $l_i$ are the parameters of Gaussian process to be found from the light curve approximation.
These length parameters describe the characteristic time scale of correlation between observations. If the value of $l_i$ is too small the approximated light curve will be over-fitted and can show unrealistic oscillations.
To prevent it we set a lower limit on $l_i$ as the maximum time interval between two neighbouring observations, but not larger than 60 days. Also, \textsc{Multivariate Gaussian Process} kernel includes 6 constants, three of which are unit variances of basis processes and three others describe their pairwise correlations.
Totally, \textsc{Multivariate Gaussian Process} has 9 parameters to be fitted.

Prior to applying the GP approximation, we prepare the data~(Fig.~\ref{workflow}). First, we transform the magnitudes given by the Open Supernova Catalog to fluxes and perform all further analysis in the flux space only. Since measurements remote in time from the maximum (mainly the upper limits or host detection) could potentially affect the GP behaviour, including the main part of light curve around maximum, for each object we take only the points in the interval $[-240, +240]$ days relative to the maximum in $r,r',R$ filter depending on the sub-sample. The Julian dates are rounded to integers. We also implement 1-day time-binning to the data.

In every bin the flux $y$ and its error $\sigma$ are derived from $n$ observations $\{y_i, \sigma_i\}$ as follows~\citep{agekian1972}:
\begin{equation}
\begin{split}
    w_i &\equiv \frac1{\sigma_i^2},\\
    w &\equiv \sum{w_i},\\
    y &= \frac{\sum{w_i\,y_i}}{w},\\
    \langle\sigma\rangle &\equiv \sqrt{\frac{\sum{w_i\,(y_i - y)^2}}{w\,(n-1)}},\\
    \sigma_w &\equiv w^{-1/2},\\
    \sigma &= \left\{ \begin{split} &\langle\sigma\rangle, &\langle\sigma\rangle > \sigma_w,\\ &\frac12 \left(\langle\sigma\rangle + \sigma_w\right), &\langle\sigma\rangle \leq \sigma_w, \end{split} \right.
\end{split}
\end{equation}
where $w_i$ is the weight of observation, $w$ is the sum of weights, $\langle\sigma\rangle$ is the mean error, $\sigma_w$ is the error of the weighted mean. If the mean error is larger than the error of the weighted mean, then observation errors are probably underestimated or the object is very variable during the considered  time interval.
Upper limits are taken into account only if there are no detections in the bin. In this case we keep the most conservative upper limit, i.e. the one with the smallest flux.

\begin{figure*}
\begin{center}
\includegraphics[scale=0.58]{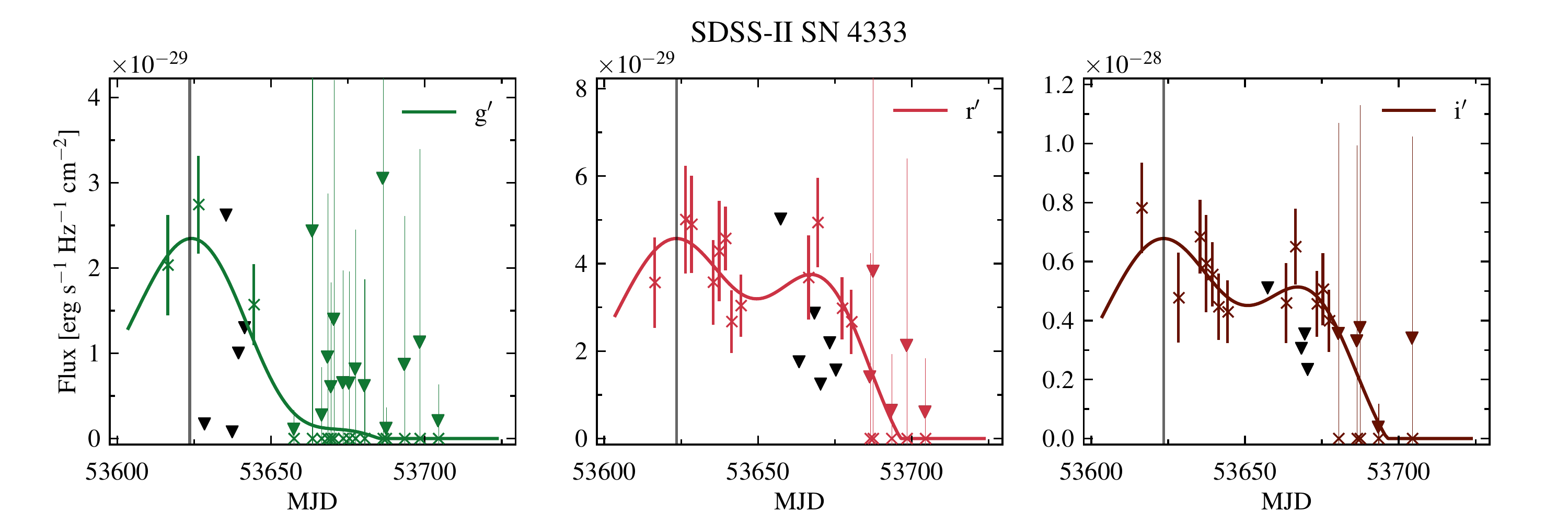}
\caption{An example of multicolour light curve explaining the pre-processing procedure. The crosses with errorbars denote the real photometric detections. Coloured triangles are the original upper limits ($y_{\rm up}$) which are either later than the latest real detection or earlier than the earliest real detection. They are transformed into the observations with $y=0\pm3y_{\rm up}$ (crosses with the thin errorbars). Black triangles are upper limits which are ignored. GP approximation of the crosses is shown by solid lines.}
\label{preproc}
\end{center}
\end{figure*}

Since for each object the OSC assembles the photometry obtained by different telescopes with different limited magnitudes, a lot of upper limits appeared in between or even simultaneously with the real detections. This could also have an undesirable impact on the Gaussian processes approximation. Therefore, for each filter we keep only those upper limits which are later than the latest real detection or earlier than the earliest real detection. Furthermore, we reassign the values of these upper limits $y_{\rm up}$: the new values are zeros with error equal to $3\times y_{\rm up}$. This is done to decrease the influence of too high upper limits on the GP approximation and to force it to vanish for very early and very late times. Some particular aspects of pre-processing are illustrated in  Fig.~\ref{preproc}.

Once the \textsc{Multivariate Gaussian Process} approximation was done, we visually inspected the resulting light curves. Those SNe with unsatisfactory approximation were removed from the sample (mainly the objects with bad photometric quality). The remaining $BRI$ approximated light curves were then transformed to $gri$ (Sect.~\ref{transformation}). 

We consider the light curves in the observer frame. Since each object has its own flux scale due to the different origin and different distance, we normalized the flux vector by its maximum value. Based on the results of this approximation, for each object we extracted the kernel parameters,  the log-likelihood of the fit, LC maximum and normalized photometry in the range of $[-20, +100]$ days with 1-day interval relative to the maximum. These values were used as features for the ML algorithm (Sect.~\ref{sec:iso_forest}).

\begin{figure*}
\begin{minipage}{0.47\linewidth}
\center{\includegraphics[width=1\linewidth]{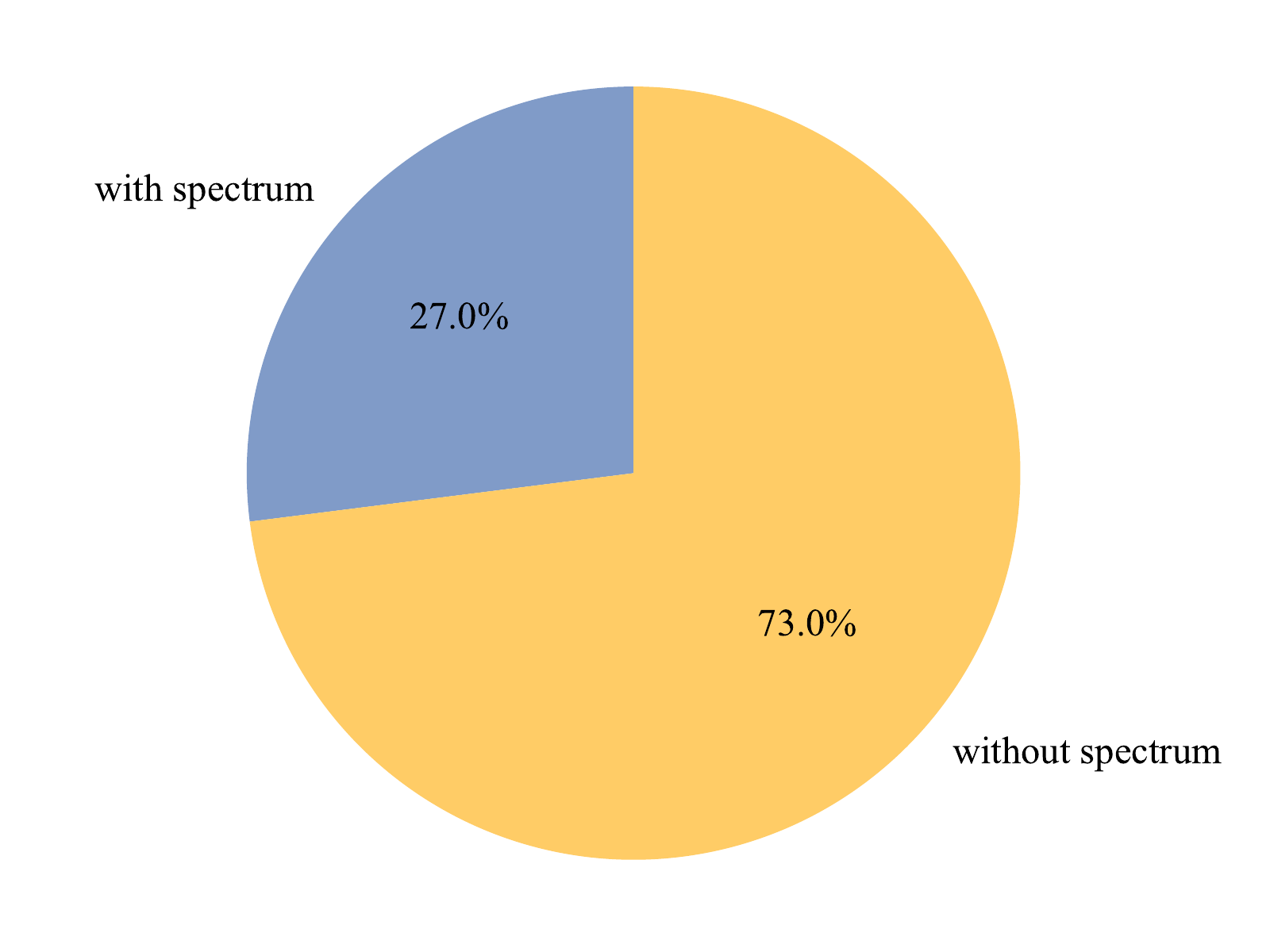} \\ (a)}
\end{minipage}
\hfill
\begin{minipage}{0.47\linewidth}
\center{\includegraphics[width=1\linewidth]{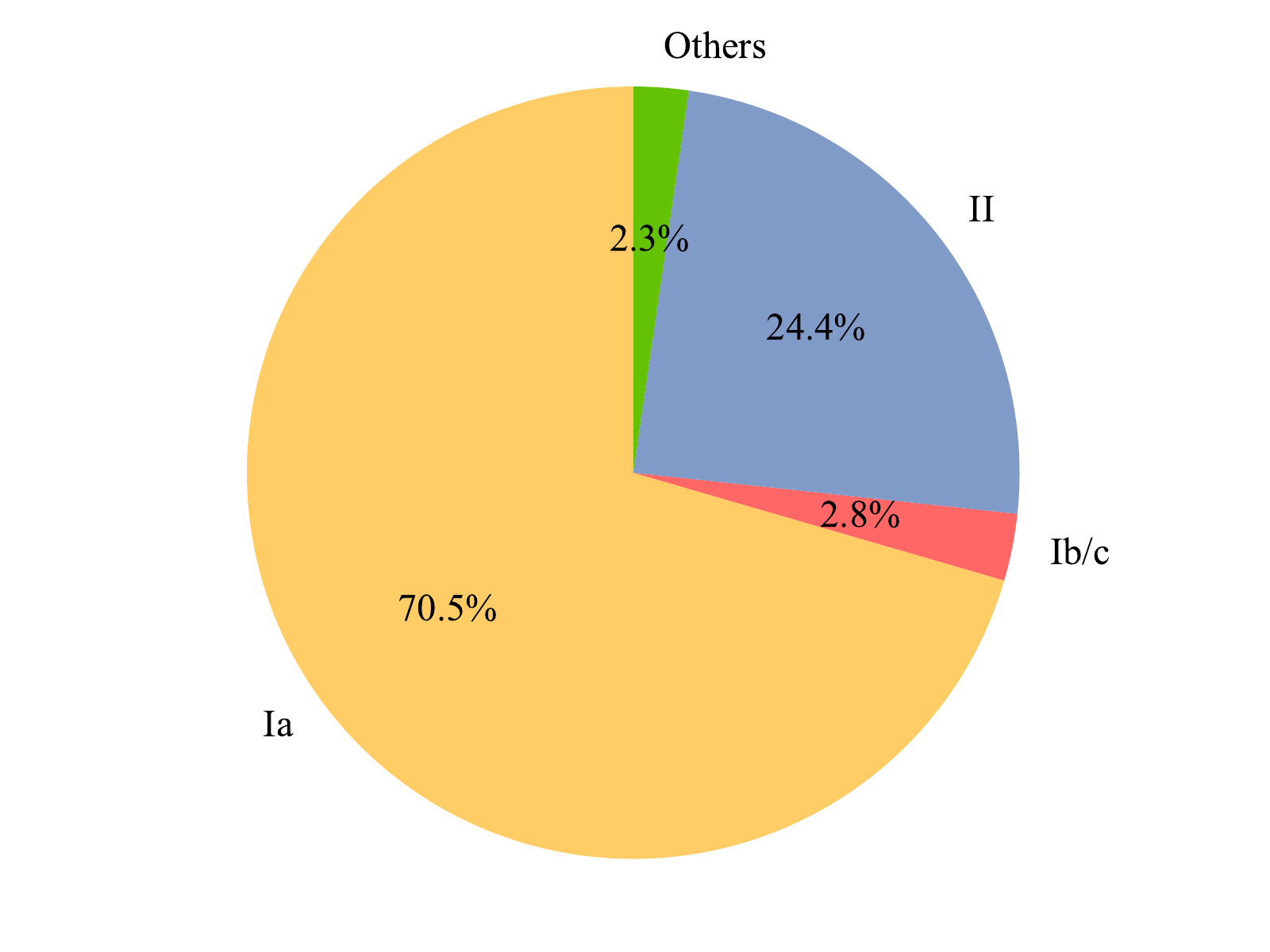} \\ (b)}
\end{minipage}
\caption{(a) Fraction of objects from our sample with at least one spectrum in the Open Supernova Catalog; (b) Distribution of these objects by the OSC types. }
\label{type_distr}
\end{figure*}

\begin{figure}
\begin{center}
\includegraphics[scale=0.5]{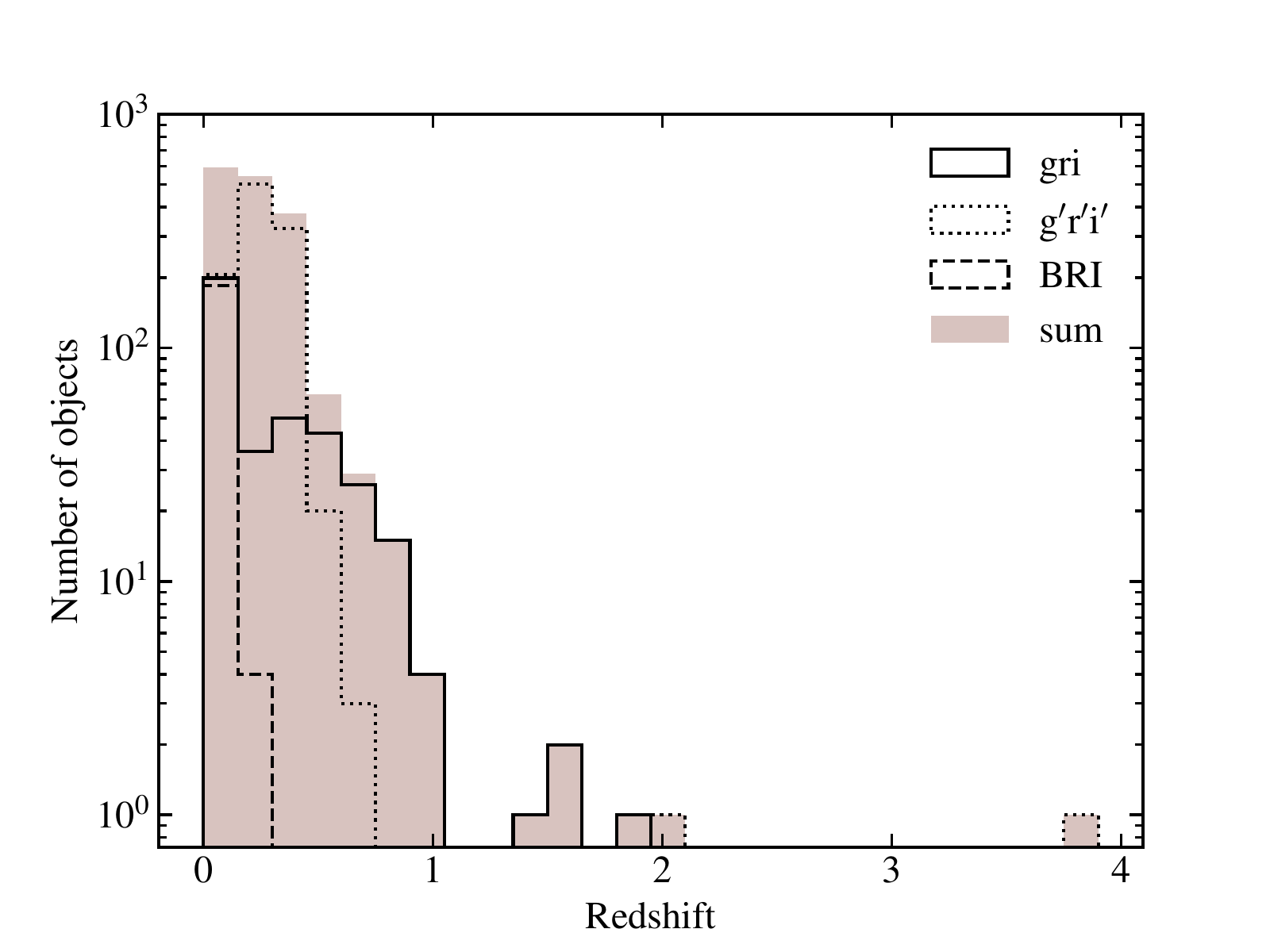}
\caption{Distribution of objects from our sample by the redshift for three sub-samples and in total. The redshift is available for 1624 objects only.}
\label{z_distr}
\end{center}
\end{figure}

\begin{figure}
\begin{center}
\includegraphics[scale=0.5]{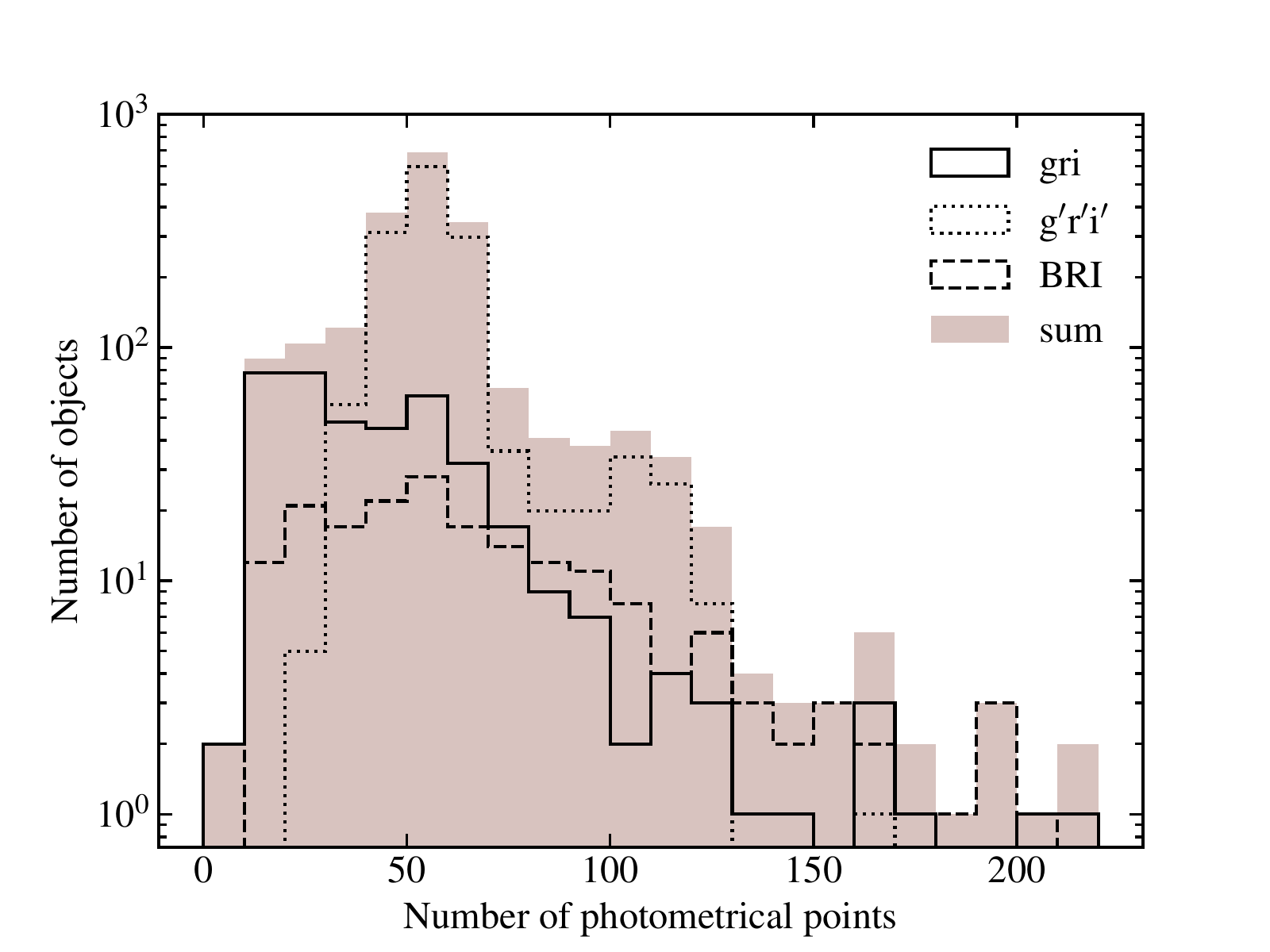}
\caption{Distribution of objects from our sample by the number of photometric points for three sub-samples and in total.}
\label{photo_distr}
\end{center}
\end{figure}

Our final sample consists of 1999 objects, $\sim$30\% of which have at least one spectrum in the OSC (see~Fig.~\ref{type_distr}). The distribution of these objects by astrophysical types is also shown in Fig.~\ref{type_distr}. The classification is extracted from the OSC without any verification, it can be photometric or based on one spectrum only. Less than 5\% of our sample have $<$20 photometric points in all three filters. The distributions of objects  by redshift and by number of photometric points for the three sub-samples are shown in Figs.~\ref{z_distr} and \ref{photo_distr}. The Fig.~\ref{z_distr} contains only 1624 objects which significantly exceeds the number of objects with the OSC spectra. The reason for such discrepancy is that, first, the OSC collects also the photometric redshifts and, second, not all spectroscopically confirmed supernovae have the spectrum available in the public domain (for details, see \citealt{2017ApJ...835...64G}).

\subsection{Dimensionality Reduction}
\label{sec:dim_red}

After the approximation procedure, each object has 374 features: $121 \times 3$ normalized fluxes, the LC flux maximum, 9 fitted parameters of the Gaussian process kernel, and the log-likelihood of the fit. 

We apply the outlier detection algorithm not only to the full data set but also to the dimensionality-reduced data. The reason for this is that the initial high dimensional feature space can be too sparse for the successful performance of the isolation forest algorithm. We applied t-SNE~\citep{maaten_hinton_2008}, a variation of the stochastic neighbour embedding method~\citep{hinton_etal2003}, for the dimensionality reduction of the data. In the t-SNE technique, a nonlinear dimensionality reduction mapping is obtained so as to keep distribution of distances between points undisturbed. This ensures that if a point is anomalous in the sense that it is distant from other points in the original data, it remains anomalous in the lower dimension space.
As a result of the dimentionality reduction, we obtain 8 separate reduced data sets corresponding to 2 to 9 t-SNE features (dimensions). 
Since t-SNE is a stochastic technique we have also taken additional precautions to ensure that the resulting outlier list does not depend on the t-SNE initial random state.

\section{Isolation Forest}
\label{sec:iso_forest}

\begin{figure}
\begin{center}
\includegraphics[width=0.9\linewidth]{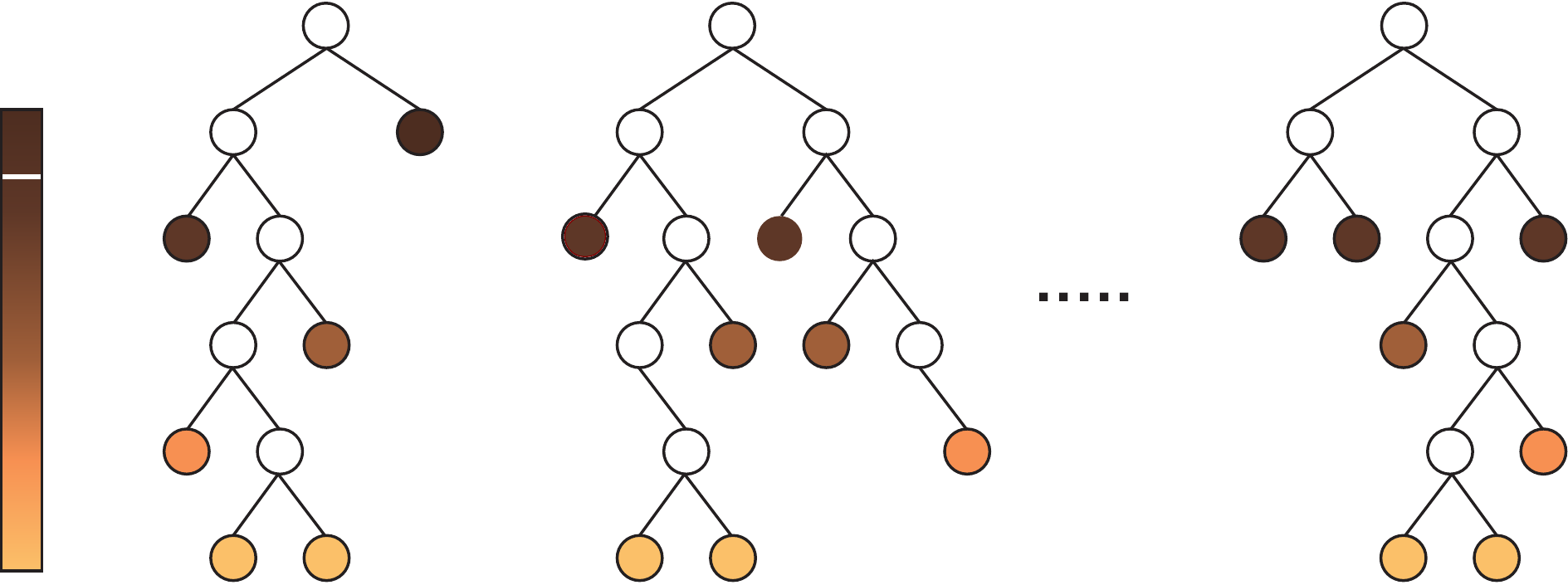}
\caption{Isolation forest structure. Forest consists of the independent decision trees. To build a branching in a tree a random feature and a random splitting are selected. The tree is built until each object of a sample is isolated in a separate leaf --- the shorter path corresponds to a higher anomaly score which is also illustrated by the colour. For each object, the measure of its normality is a function of the depths of the leaves into which it is isolated.} 
\label{cartoon_isoforest}
\end{center}
\end{figure}

Isolation forest~\citep{liu2008isolation,Liu:2012:IAD:2133360.2133363} is an ensemble of random isolation trees. Each isolation tree is a space partitioning tree similar to the widely-known Kd-tree~\citep{Bentley:1975:MBS:361002.361007}. However, in contrast to the Kd-tree, a space coordinate (a feature) and a split value are selected at random for every node of the isolation tree. This algorithm leads to an unbalanced tree unsuitable for efficient spatial search. However, the tree has the following important property: a path distance between the root and the leaf is shorter on average for points distant from "normal" data. This allows us to construct enough random trees to estimate average root-leaf path distance for every data sample that we have, and then rank the data samples based on the path length. The anomaly score, defined in a range  $[0, 1]$, is assigned to each object (see Eq.~2 in~\citealt{liu2008isolation}). Then, objects with the highest anomaly score --- outliers --- are selected according to the contamination level which is a hyper-parameter of the algorithm.  The isolation forest algorithm is illustrated in Fig.~\ref{cartoon_isoforest}.

We run the isolation forest algorithm on 10 data sets obtained using the same photometric data~(Fig.~\ref{workflow}): 
\begin{enumerate}[label=\Alph*)]
    \item data set of 364 photometric characteristics ($121 \times 3$ normalized fluxes, the LC flux maximum),
    \item data set of 10 parameters of the Gaussian process (9 fitted parameters of the kernel, the log-likelihood of the fit),
    \item 8 data sets obtained by reducing 374 features to 2--9 t-SNE  dimensions~(Sect.~\ref{sec:dim_red}).
\end{enumerate}

\begin{figure}
\begin{center}
\includegraphics[scale=0.5]{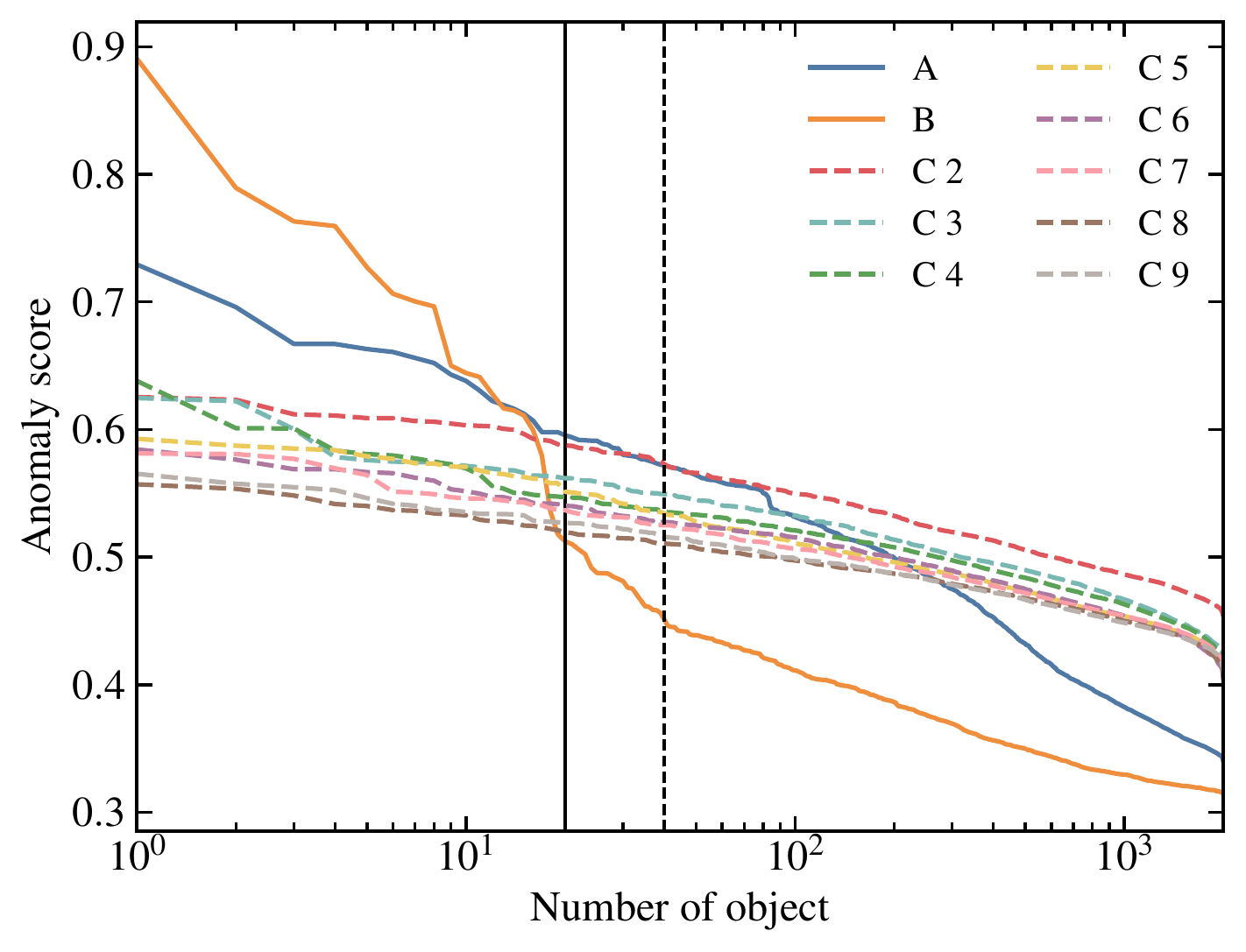}
\caption{Distribution of objects by anomaly score in 10 data sets described in Sect.~\ref{sec:iso_forest}, C~2 -- C~9 denote C data sets with 2--9 t-SNE dimensions. In each data set objects are ordered by score. Black solid and dashed lines denote 1\% and 2\% contamination level of outliers, respectively.} 
\label{score}
\end{center}
\end{figure}

For each data set we obtained a list of outliers. Contamination levels were set to 1\% (20 objects with highest anomaly score) for data sets A and B. For all data sets in case C we considered 2\% contamination (40 objects with highest anomaly score). This larger contamination was chosen to take into account the influence of the dimensionality reduction step in the final data configuration. Given different representations of the data and the stochastic nature of the isolation forest algorithm, the same object can be assigned a different anomaly score depending on how many t-SNE dimensions are used. Thus, only those objects which were listed within the 2\% contamination in at least 2 of the data sets in case C are included in Table~\ref{outliers_table} and subjected to further astrophysical analysis. 
The distribution of objects in each of 10 data sets by anomaly score is presented in Fig.~\ref{score}.

\begin{figure}
\centering
\includegraphics[trim={0.0cm 0.35cm 0 0},clip,scale=0.46]{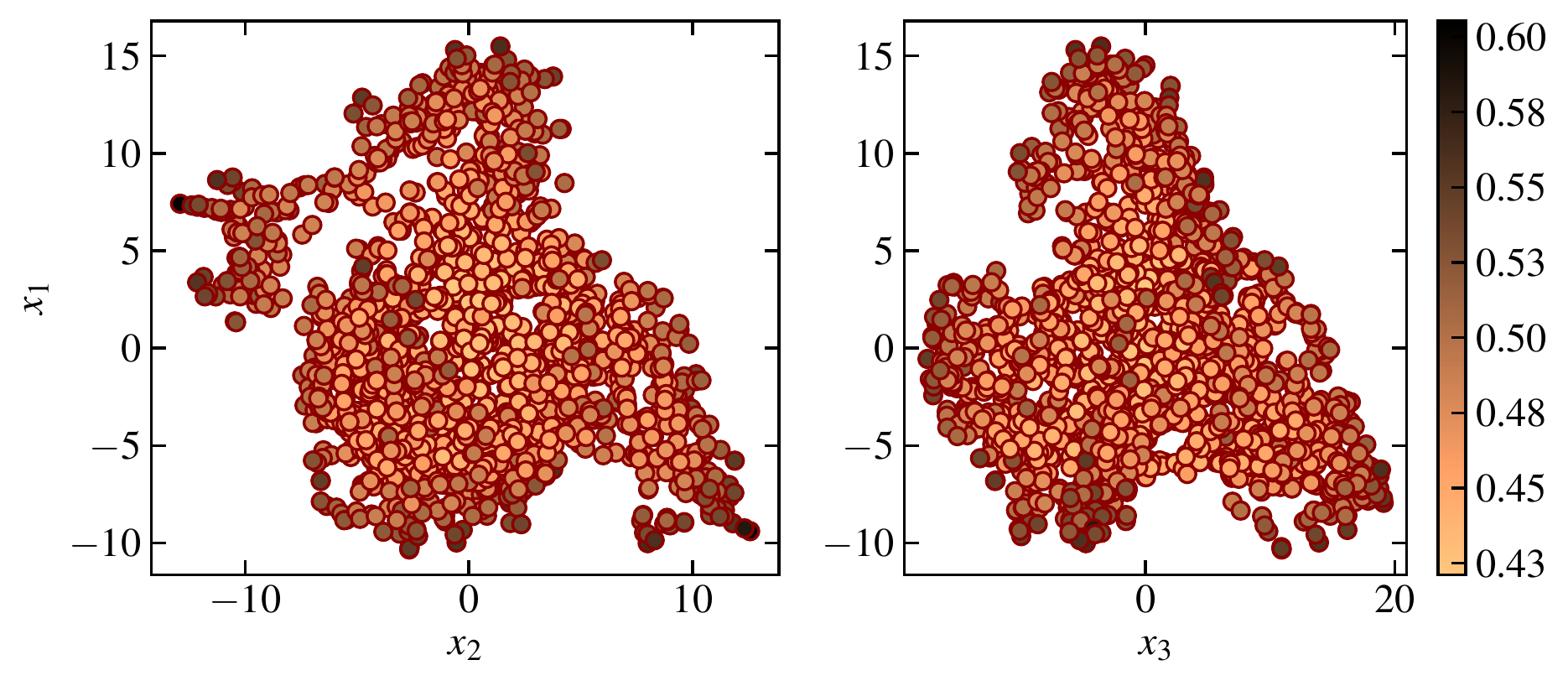}
\caption{Three-dimensional t-SNE reduced data after application of the isolation forest algorithm. Each point represents a supernova light curve from the data set projected into the three-dimensional space with the coordinates $(x_1, x_2, x_3)$. The intensity of the colour indicates the anomaly score for each object as estimated by the isolation forest algorithm. A darker color corresponds to the objects with higher anomaly scores.}
\label{tsne_isoforest}
\end{figure}

An example of the isolation forest algorithm applied to the three-dimensional reduced data set is shown in Fig.~\ref{tsne_isoforest}.

\section{Results}
\label{sec:results}
Applying the unsupervised learning to the photometric data extracted from the Open Supernova Catalog we found $\sim$100 outliers among a total of 1999 objects~(Fig.~\ref{workflow}). However, not all of them are necessary anomalies. That is why we also subject the outliers to the careful
astrophysical analysis. Using publicly available sources, we collected  information about each outlier and determined to which kind of astrophysical objects it belongs --- given the information we could gather. Among the detected outliers there are few known cases of miss-classifications, representatives of rare classes of SNe (e.g., superluminous supernovae, 91T-like SNe~Ia) and highly reddened objects. We also found that 16 anomalies classified as supernovae by~\cite{Sako2018}, are likely to be quasars or stars.

Light curves with GP approximation for all 1999 objects can be found at \href{http://snad.space/osc/}{http://snad.space/osc/} and those who considered anomalous according to the criteria described in the previous section are listed in Table~\ref{outliers_table}. Names and equatorial coordinates of outliers are shown in Columns 1-3; types in Column~4. CMB redshifts are presented in Column~5. Columns 6-8 contain the names and equatorial coordinates of the corresponding host galaxies. Host morphological types are displayed  in Column~9. Columns 10 and 11 contain the separation between center of the host and object in angular seconds and kiloparsecs, respectively (to calculate the angular diameter distance we use a flat $\Lambda$CDM cosmology with $H_0 = 70$~km~s$^{-1}$~Mpc$^{-1}$, $\Omega_\Lambda = 0.7$). We give our comments and short description of each object in Column~12. References are in Column~13.  
The most interesting of these objects are described below. 

\subsection{Peculiar SNe~Ia}

Type Ia supernova is an explosion of a carbon-oxygen white dwarf that exceeds the Chandrasekhar limit either by matter accretion from a companion star or by merging with another white dwarf~\citep{1973ApJ...186.1007W,1984ApJS...54..335I,1984ApJ...277..355W}. SNe~Ia are used as universal distance ladder since their luminosity at maximum light is approximately the same~\citep{Perlmutter99,Riess98}. However, the class of SNe~Ia is not homogeneous, for example, 91T-like supernovae are on average 0.2--0.3~mag more luminous than normal SNe~Ia, have broader
LCs, and different early spectrum evolution~\citep{1992ApJ...384L..15F,2012AJ....143..126B}; 91bg-like supernovae are subluminous and fast-declining~\citep{1992AJ....104.1543F};  peculiar SNe~Iax are spectroscopically similar to SNe~Ia, but have lower maximum-light velocities and typically lower peak magnitudes~\citep{2013ApJ...767...57F}.
The presence of non-classical SNe~Ia in cosmological samples may introduce a systematic bias and affect the cosmological analysis~(e.g.,~\citealt{2012ApJ...757...12S}).

\subsubsection{SN2002bj}

\begin{figure*}
\begin{center}
\includegraphics[scale=0.58]{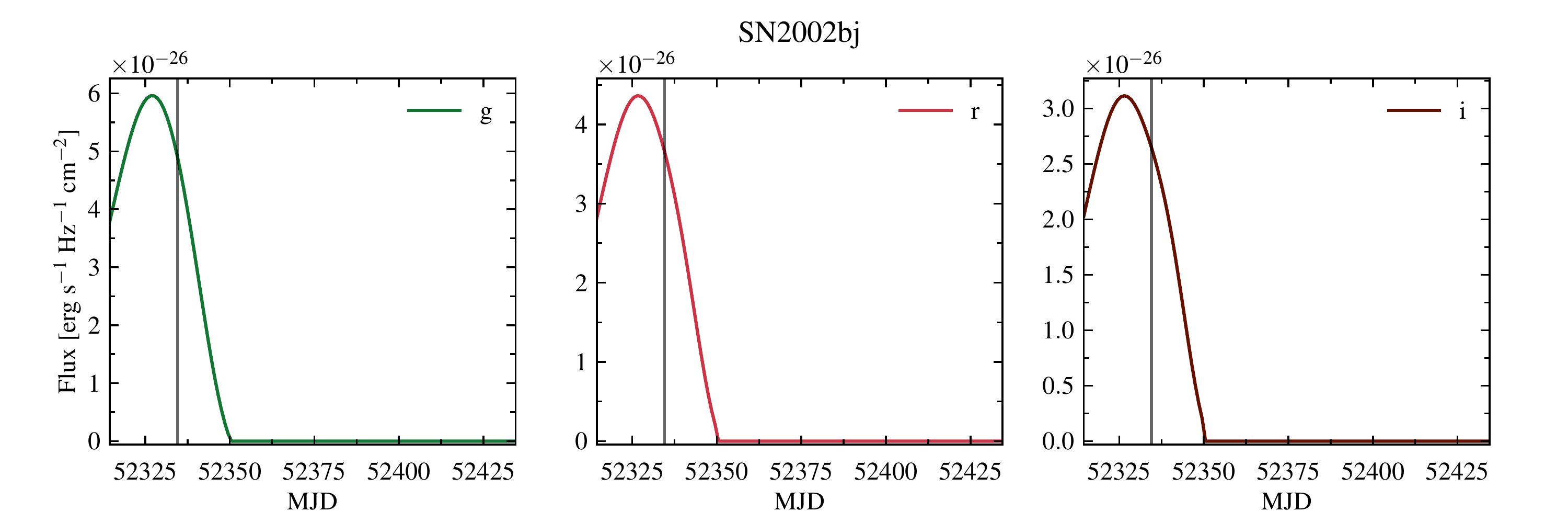}
\caption{Light curves in $gri$ filters of peculiar SN~Ia 2002bj~\citep{2010Sci...327...58P}. Solid lines are the results of our approximation by \textsc{Multivariate Gaussian Process}. The LCs in $gri$ filters are obtained from the Bessel's $BRI$  by filter transformation~(Sect.~\ref{transformation}), thus the observations are absent on the plot. The vertical line denotes the moment of maximum in $R$ filter.}
\label{SN2002bj}
\end{center}
\end{figure*}

SN2002bj was discovered in NGC~1821 on unfiltered CCD frames taken with the Puckett Observatory 0.60-m automated patrol telescope on 2002 February 28.06 and March 1.05~UT, and on unfiltered CCD LOTOSS images taken with the 0.8-m Katzman Automatic Imaging Telescope on February 28.2 and March 1.2~UT~\citep{2002IAUC.7839....1P}. This supernova was a first representative of rapidly evolving events (Fig.~\ref{SN2002bj}). Its light curve has a rise time of $<$7 days followed by a decline of $0.25$~mag day$^{-1}$ in $B$ band and reaches a peak intrinsic brightness greater than $-18$ mag~\citep{2010Sci...327...58P}. The spectra are similar to that of a SN~Ia but show the presence of helium and carbon lines. The analysis of archive data after the discovery of this object and the subsequent observations revealed other bright, fast-evolving supernovae, e.g., SN1885A, SN1939B, SN2010X, SN2015U~\citep{2010ApJ...723L..98K,Perets2011, 2016MNRAS.461.3057S}. These objects can be produced by the detonation of a helium shell on a white dwarf, ejecting a small envelope of material~\citep{2010Sci...327...58P}.

\subsubsection{SN2013cv}

\begin{figure*}
\begin{center}
\includegraphics[scale=0.58]{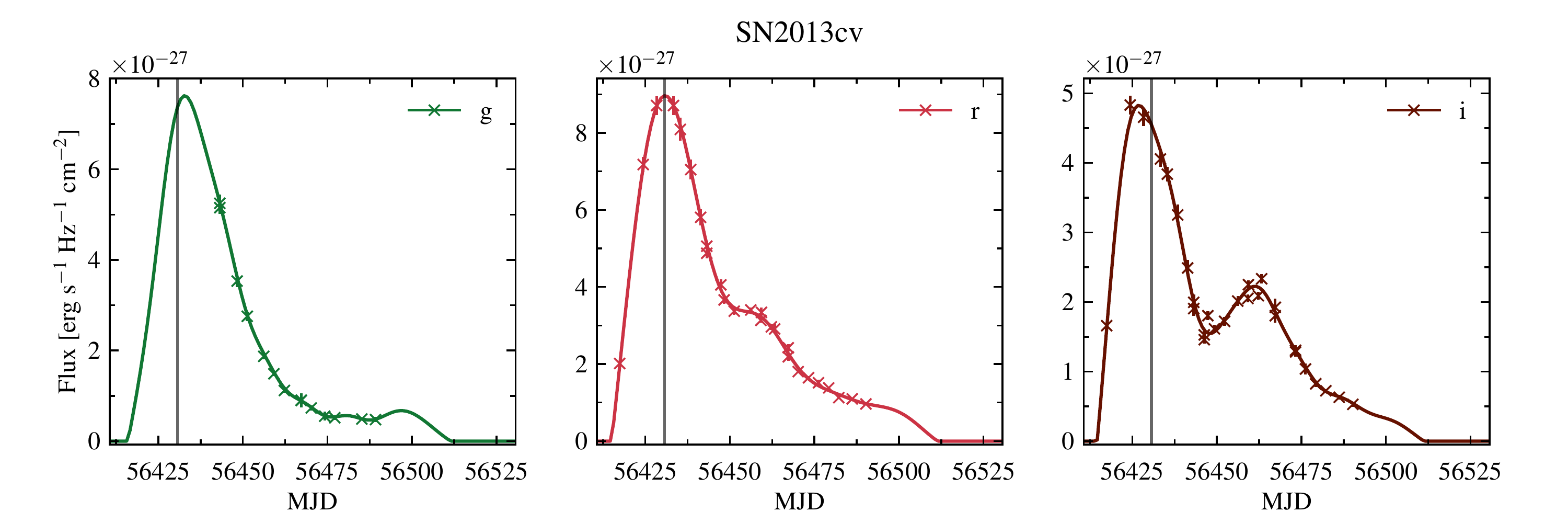}
\caption{Light curves in $gri$ filters of peculiar SN~Ia 2013cv~\citep{0004-637X-823-2-147,2012PASP..124..668Y}. Solid lines are the results of our approximation by \textsc{Multivariate Gaussian Process}. The vertical line denotes the moment of maximum in $r$ filter.}
\label{SN2013cv}
\end{center}
\end{figure*}

 SN2013cv was independently discovered by \cite{2013CBET.3543....1Z} and iPTF~\citep{2009PASP..121.1395L} on 2013 May 1.44 UT, see~Fig.~\ref{SN2013cv}. This peculiar supernova has large peak optical and UV luminosity and show an absence of iron absorption lines in the early spectra. \cite{0004-637X-823-2-147} suggests that SN2013cv is an intermediate case between the normal and super-Chandrasekhar events.

\subsubsection{SN2016bln}

\begin{figure*}
\begin{center}
\includegraphics[scale=0.58]{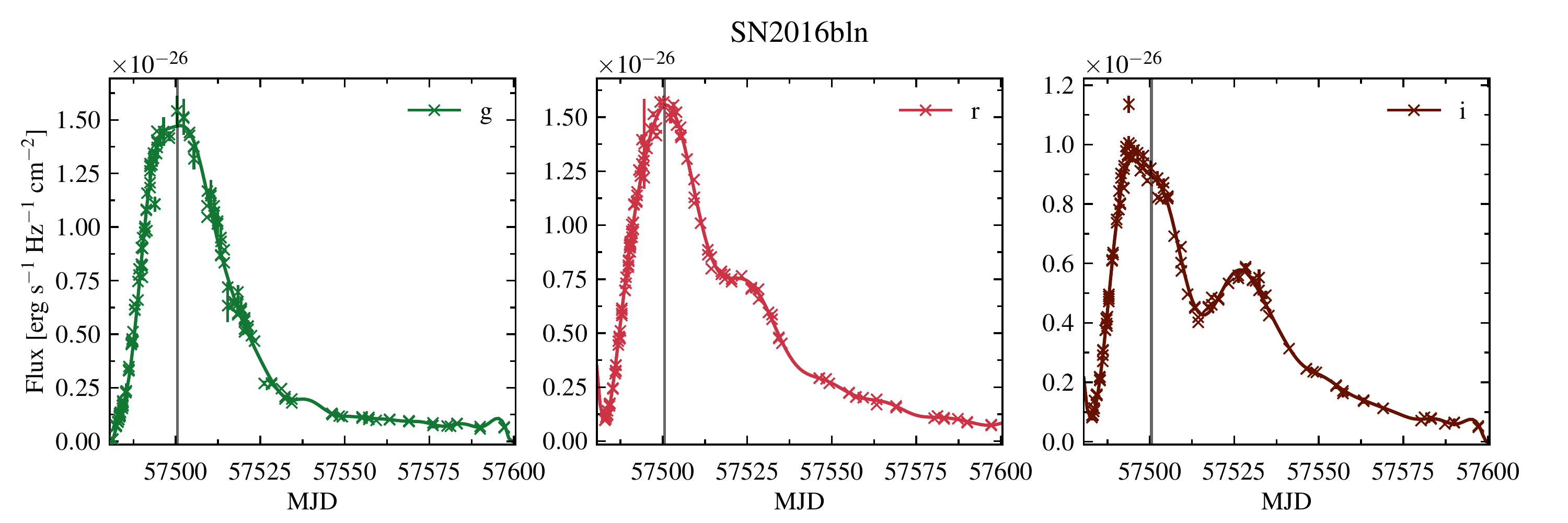}
\caption{Light curves in $gri$ filters of 91T-like SN~Ia 2016bln~\citep{2018ApJ...852..100M}. Solid lines are the results of our approximation by \textsc{Multivariate Gaussian Process}. The vertical line denotes the moment of maximum in $r$ filter.}
\label{SN2016bln}
\end{center}
\end{figure*}

SN2016bln/iPTF16abc discovered by the iPTF on 2016 April l3.36  UT~\citep{2016ATel.8907....1M,2016ATel.8909....1C} and classified by our code as outlier, belongs to the 91T-like SNe~Ia subtype (see~Fig.~\ref{SN2016bln}). The transitional and nebular spectrum of SN2016bln appear similar to the normal SN2011fe as well as to over-luminous SNe 1991T and 1999aa~\citep{2018MNRAS.480.1445D}. Early-time  observations  show  a  peculiar  rise  time,  non-evolving blue colour, and  unusual strong $\rm C~II$ absorption. These features can be explained by the ejecta interaction with nearby, unbound material or/and significant $^{56}$Ni mixing within the SN ejecta~\citep{2018ApJ...852..100M}.

\subsection{Peculiar SNe~II}

Type II supernovae arise from the core collapse of massive stars at the final stage of their evolution. The radius of these stars can be several hundred times greater than the solar radius, and their extremely tenuous envelopes contain large amounts of hydrogen. That is why hydrogen lines are the most prominent in the spectra of SNe~II. Based on the shape of light curves Type II supernovae have historically been divided into the Type~IIL (linear) and Type~IIP (plateau) subtypes, however the following studies revealed a continuity in light curve slopes of Type II SNe~\citep{Anderson2014a, Sanders2015}.

\subsubsection{SN2013ej}

\begin{figure*}
\begin{center}
\includegraphics[scale=0.58]{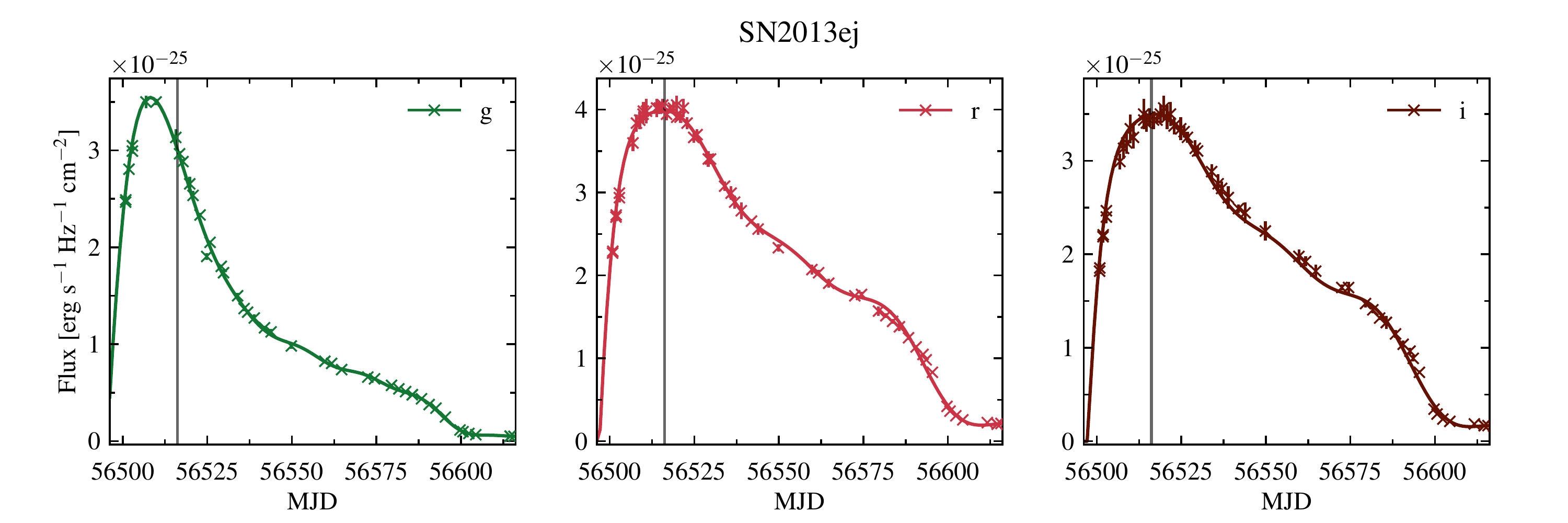}
\caption{Light curves in $gri$ filters of peculiar SN~II~2013ej~\citep{2016MNRAS.461.2003Y}. Solid lines are the results of our approximation by \textsc{Multivariate Gaussian Process}. The vertical line denotes the moment of maximum in $r$ filter.}
\label{SN2013ej}
\end{center}
\end{figure*}

Light curve of SN2013ej, discovered by the Lick Observatory Supernova Search on 2013 July 25.45~UT~\citep{2013CBET.3606....1K}, appears intermediate between those of Type IIP and IIL supernovae (see~Fig.~\ref{SN2013ej}). The event has a higher peak luminosity, a faster post-peak decline, and a shorter plateau phase compared to the normal Type IIP SN 1999em. The radioactive $^{56}$Ni mass is 0.02 ${M}_{\odot }$, which is significantly lower than for typical SNe IIP~\citep{2015ApJ...807...59H}. The source exhibits signs of substantial geometric asphericity, X-rays from persistent interaction with circumstellar material (CSM), thermal emission from warm dust~\citep{2017ApJ...834..118M}.

\subsubsection{SN2016ija}

\begin{figure*}
\begin{center}
\includegraphics[scale=0.58]{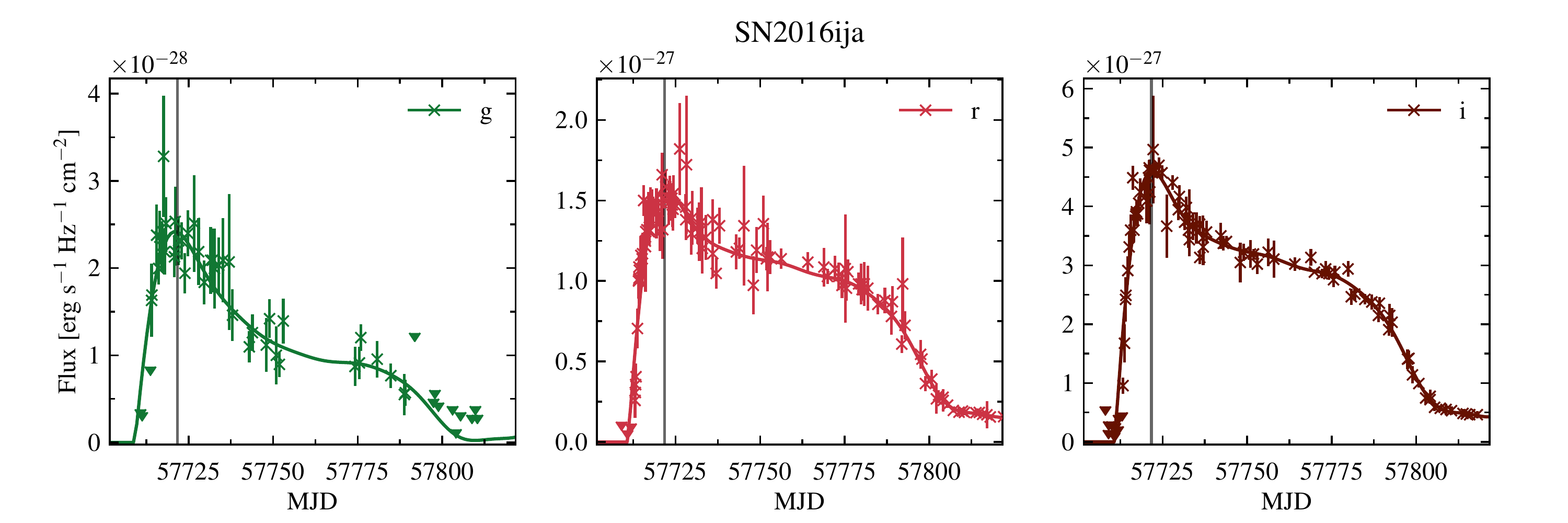}
\caption{Light curves in $gri$ filters of peculiar SN~II~2016ija~\citep{2018ApJ...853...62T}. Solid lines are the results of our approximation by \textsc{Multivariate Gaussian Process}. The vertical line denotes the moment of maximum in $r$ filter.}
\label{SN2016ija}
\end{center}
\end{figure*}

This supernova was discovered on 2016 November 21.19~UT (\citealt{2016ATel.9782....1T}, see Fig.~\ref{SN2016ija}) during one-day cadence SN search for very young  transients in the nearby Universe (DLT40). Using SNID~\citep{SNID}, it was first suggested to be an early time 91T-like SN Ia with few features and red continuum. It has been also associated to the outburst in an obscured luminous blue variable, an intermediate luminosity red transient or a luminous red nova~\citep{2016ATel.9787....1B}. The subsequent spectroscopic follow-up revealed broad $\rm H_{\alpha}$ and calcium features, leading to a classification as a highly extinguished Type~II supernova. The colour excess from the host galaxy NGC~1532 is $E(B-V)_{\rm host} = 1.95\pm0.15$~mag~\citep{2018ApJ...853...62T}. Moreover, SN2016ija is brighter than usual SNe~II (see fig.~6 of~\citealt{2018ApJ...853...62T}).

\subsection{Superluminous SNe}

Superluminous SNe are supernovae with an absolute peak magnitude $M < -21$~mag in any band. According to~\cite{2012Sci...337..927G} SLSN can be divided into three broad classes: SLSN-I without hydrogen in their spectra, hydrogen-rich SLSN-II that often show signs of interaction with CSM, and finally, SLSN-R, a rare class of hydrogen-poor events with slowly evolving LCs, powered by the radioactive decay of $^{56}$Ni. SLSN-R are suspected to be pair-instability supernovae: the deaths of stars with initial masses between 140 and 260 solar masses.

In our outlier list in Table~\ref{outliers_table} there are four SLSN: SDSS-II~SN~17789, SN2015bn, PTF10aagc, SN2213-1745.

\subsubsection{SN2213-1745}

\begin{figure*}
\begin{center}
\includegraphics[scale=0.58]{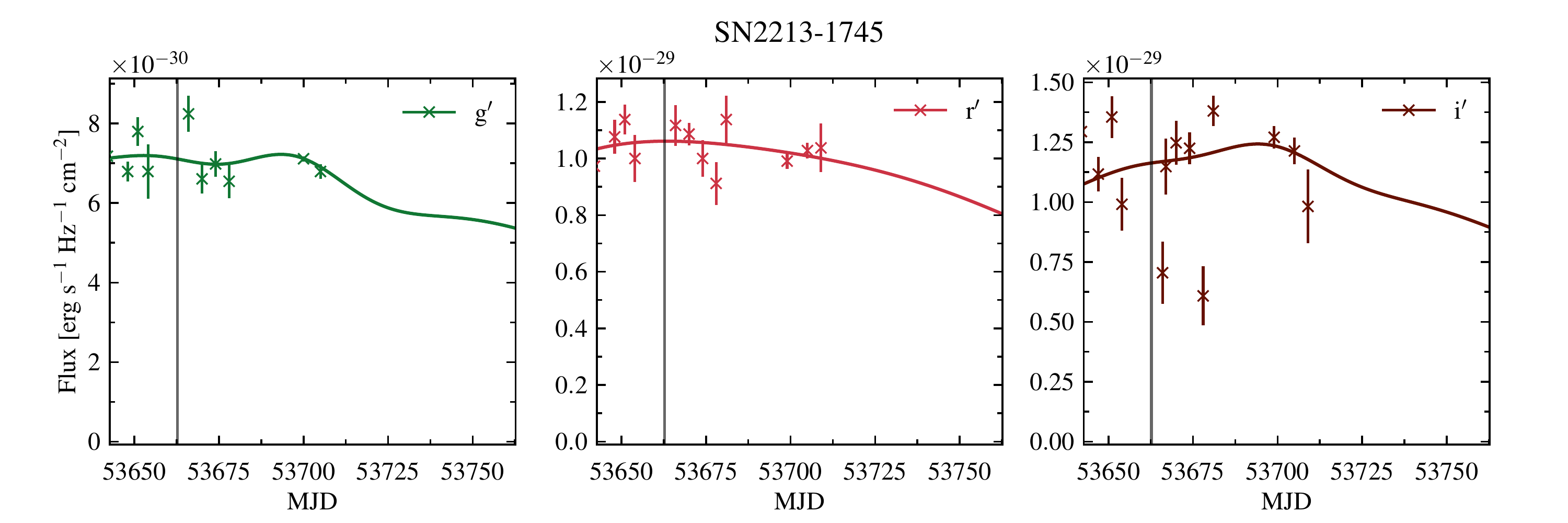}
\caption{Light curves in $g'r'i'$ filters of superluminous supernova SN2213-1745~\citep{2012Natur.491..228C}. Solid lines are the results of our approximation by \textsc{Multivariate Gaussian Process}. The vertical line denotes the moment of maximum in $r'$ filter.}
\label{SN2213-1745}
\end{center}
\end{figure*}

SN2213-1745 was discovered at  $z=2.046$ by the Canada-France-Hawaii Telescope Legacy Survey (Fig.~\ref{SN2213-1745}). It belongs to the SLSN-R events. \cite{2012Natur.491..228C} suggested that SN 2213-1745 may be powered by the radiative decay of a 4--7~$M_\odot$ of synthesised $^{56}$Ni, and implied a progenitor with an estimated initial mass of $\sim$250~$M_\odot$.  

\subsubsection{PTF10aagc}

\begin{figure*}
\begin{center}
\includegraphics[scale=0.58]{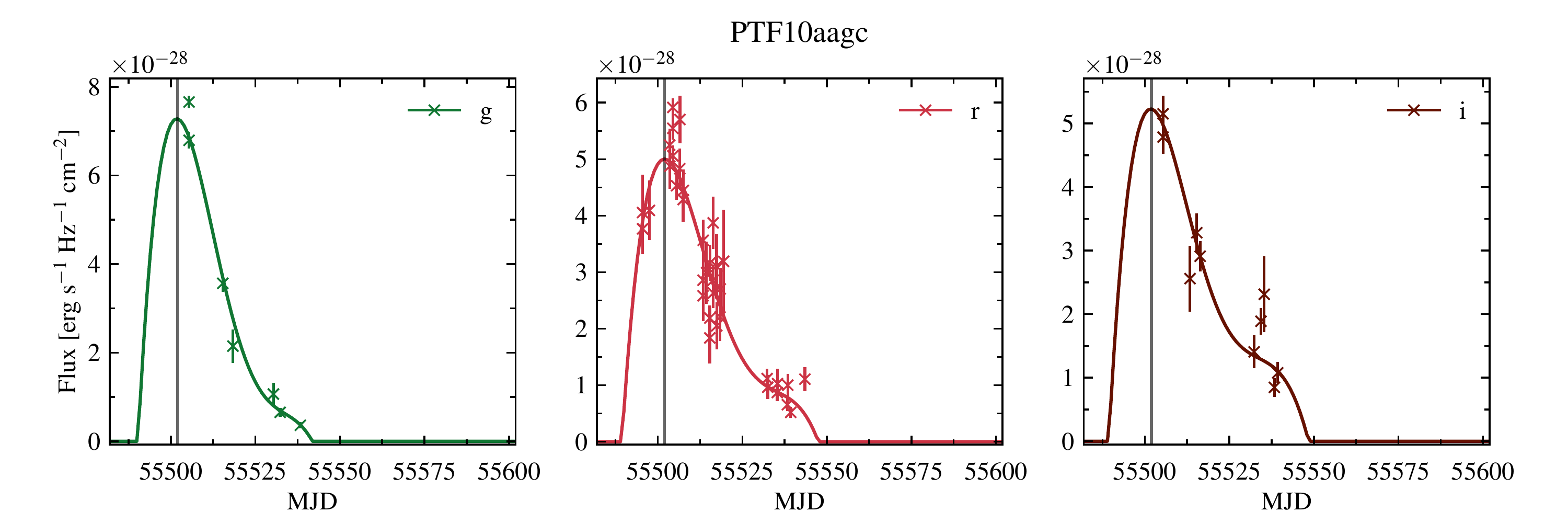}
\caption{Light curves in $gri$ filters of superluminous supernova PTF10aagc~\citep{2018ApJ...860..100D}. Solid lines are the results of our approximation by \textsc{Multivariate Gaussian Process}. The vertical line denotes the moment of maximum in $r$ filter.}
\label{PTF10aagc}
\end{center}
\end{figure*}

The high peak luminosity ($L_{\rm bol, peak} = 10^{43.7}$ erg~s$^{-1}$) and the absence of hydrogen lines in early spectrum allowed to attribute PTF10aagc to SLSN-I~(\citealt{2018ApJ...860..100D}, see Fig.~\ref{PTF10aagc}). However, the latter spectra revealed a broad $\rm H_\alpha$ and the corresponding weak, but detected $\rm H_\beta$~\citep{2015ApJ...814..108Y}.  This particularity makes PTF10aagc clearly distinct from others SLSN-I. Such spectral behaviour can be explained by interaction between SLSN-I ejecta and a H-rich circumstellar material at late times~\citep{2015ApJ...814..108Y}.
The host of PTF10aagc is bright and shows clear morphological structure suggesting a possible ongoing merger~\citep{2016ApJ...830...13P}. 

\subsection{Misclassified objects}

\subsubsection{SN2006kg}

\begin{figure*}
\begin{center}
\includegraphics[scale=0.58]{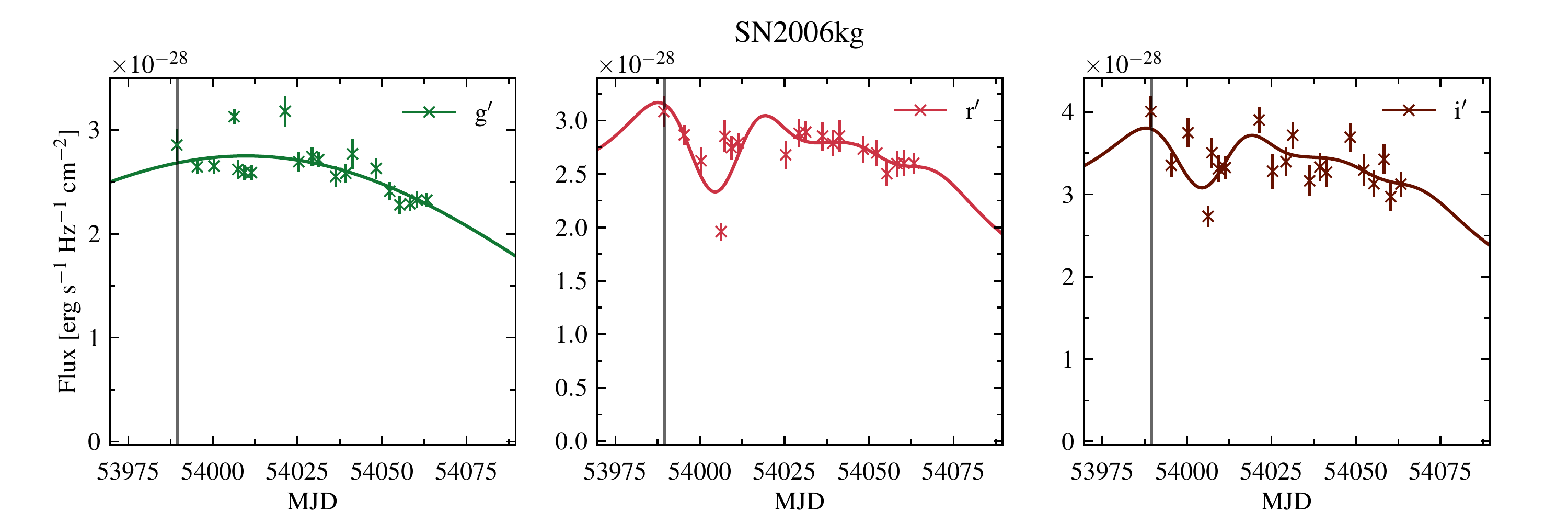}
\caption{Light curves in $g'r'i'$ filters of active galactic nucleus SN2006kg~\citep{Sako2018}. Solid lines are the results of our approximation by \textsc{Multivariate Gaussian Process}. The vertical line denotes the moment of maximum in $r'$ filter.}
\label{SN2006kg}
\end{center}
\end{figure*}

SN2006kg was first classified as a possible Type II SN~(\citealt{2006CBET..688....1B}, see~Fig.~\ref{SN2006kg}). It is also appeared as Type II spectroscopically confirmed supernova in table~6 of~\cite{2008AJ....135..348S}. However, further analysis of 3.6-m New Technology Telescope spectrum revealed that SN2006kg is an active galactic nucleus~\citep{2011AA...526A..28O,Sako2018}. It is interesting that SN2006kg continues to appear as supernova in host studies~\citep{2012A&A...544A..81H} and was even in a set of 12 well-observed events that were used as Type~II supernova templates~\citep{2014PASJ...66...49O}. The object is not in the WISE AGN Catalog~\citep{2018ApJS..234...23A} that consists of $>$20 millions AGN candidates. 

\subsubsection{Gaia16aye}

\begin{figure*}
\begin{center}
\includegraphics[scale=0.58]{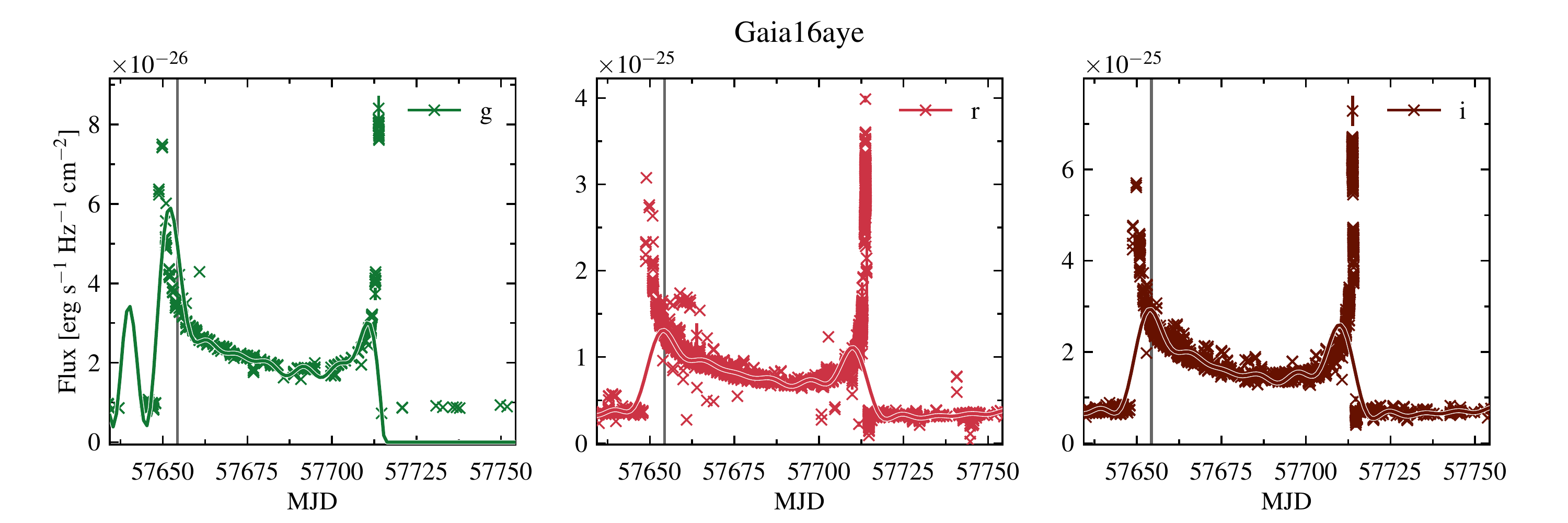}
\caption{Light curves in $gri$ filters of binary microlensing event Gaia16aye (\protect\href{http://gsaweb.ast.cam.ac.uk/alerts/alert/Gaia16aye/followup}{http://gsaweb.ast.cam.ac.uk/alerts/alert/Gaia16aye/followup}). Solid lines are the results of our approximation by \textsc{Multivariate Gaussian Process}. The vertical line denotes the moment of maximum in $r$ filter.}
\label{Gaia16aye}
\end{center}
\end{figure*}

Gaia16aye~\citep{2016ATel.9376....1B} is an object with the most non-SN-like  behavior among our set of outliers (Fig.~\ref{Gaia16aye}). In~\cite{2016ATel.9507....1W} it was reported that Gaia16aye is a binary microlensing event --- gravitational microlensing of binary systems --- the first ever discovered towards the Galactic Plane.

\subsubsection{Possible misclassified objects}

Our analysis also reveals that 16 objects classified as pSN by ~\citealt{Sako2018}, where a prefix "p" indicates a purely photometric type, are likely to be stars or quasars. First, we do not find any signature of supernovae on the corresponding multicolour light curves. Then, according to SDSS DR15\footnote{\href{http://skyserver.sdss.org/dr15/en/tools/explore/summary.aspx}{http://skyserver.sdss.org/dr15/en/tools/explore/summary.aspx}} type of SDSS-II~SN~5314, SDSS-II~SN~14170, SDSS-II~SN~15565, SDSS-II~SN~13725, SDSS-II~SN~13741, SDSS-II~SN~19699, SDSS-II~SN~18266, SDSS-II~SN~4226, SDSS-II~SN~2809, SDSS-II~SN~6992 is denoted as \textbf{STAR}. Moreover, all these objects can be found in Pan-STARRS Catalog with Pan-STARRS magnitudes equal or even brighter than those on the corresponding light curves.

The other objects SDSS-II~SN~1706, SDSS-II~SN~17756, SDSS-II~SN~17339, SDSS-II~SN~17509, SDSS-II~SN~4652, SDSS-II~SN~19395 have a BOSS~\citep{2013AJ....146...32S} spectrum with class "QSO" and have high redshifts (see Table~\ref{outliers_table})

\section{Conclusions}
\label{sec:conclusions}

The development of large sky surveys has led to a discovery of a huge number of supernovae and supernova candidates. Among the SNe discovered every year, only 10\% have spectroscopic confirmation. The amount of astronomical data increases dramatically with time and is already beyond human capabilities. The astronomical community already has dozens of thousands of SN candidates, and LSST survey~\citep{2009arXiv0912.0201L} will discover over ten million supernovae in the forthcoming decade. Only a small fraction of them will receive a spectroscopic confirmation. This motivates a considerable effort in photometric classification of supernovae by types using machine learning algorithms. There is, however, another aspect of the problem: any large photometric SN database would suffer from the non-SN contamination (novae, kilonovae, GRB afterglows, AGNs, etc.). Moreover, the database will inevitably contain the astronomical objects with unusual physical properties --- anomalies. Finding such objects and studying them in detail is very important and constitutes the main goal of this paper. 

The analysis presented here is based on the photometric data extracted from the Open Supernova Catalog~\citep{2017ApJ...835...64G}. The use of real data allows us to reveal a lot of caveats in observations at the pre-pocessing stage -- many of which are not normally present in the simulated data. After pre-processing, we obtain 1999 SNe with light curves either in $gri$ or in $g'r'i'$ or in $BRI$ filters approximated by Gaussian processes. We consider 10 different data sets: one that includes the approximated photometric observations (A), another with the parameters of Gaussian process only (B) and 8 data sets were the information in  GP photometry and GP parameters were summarised via dimensionality reduction using t-SNE (dimension varying from 2 to 9, case C). 

We apply the isolation forest algorithm to all data sets, considering a 1\% contamination for cases A/B and 2\% contamination for all data sets in case C. We visually checked all objects identified in cases A and B. We also checked all the objects which were identified as anomalous in at least 2 of the data sets in case C. 
As a result, we find $\sim$100 outliers, 40 from cases A/B and 60 which were identified in at least two data sets of case C. Among these,  19  objects were identified by both strategies, with and without dimensionality reduction. Our final validation analysis resulting in 81  objects which were carefully studied with the use of publicly available information. Among these there are four superluminous supernovae (SDSS-II~SN~17789, SN2015bn, PTF10aagc, SN2213-1745), non-classical Type~Ia SNe (91T-like SNe 2016bln, PS15cfn, SNLS-03D1cm; peculiar SN2002bj and SN2013cv), two unusual Type~II SNe which anomalous multicolour light curve behaviour can be due to the environment (SN2013ej, SN2016ija), one AGN -- SN2006kg, and one binary microlensing event Gaia16aye. We also find that 16 anomalies classified as supernovae by~\cite{Sako2018}, which are likely to be stars (SDSS-II~SN~5314, SDSS-II~SN~14170, SDSS-II~SN~15565, SDSS-II~SN~13725, SDSS-II~SN~13741, SDSS-II~SN~19699, SDSS-II~SN~18266, SDSS-II~SN~4226, SDSS-II~SN~2809, SDSS-II~SN~6992) or quasars (SDSS-II~SN~1706, SDSS-II~SN~17756, SDSS-II~SN~17339, SDSS-II~SN~17509, SDSS-II~SN~4652, SDSS-II~SN~19395). However, without careful spectral analysis it is difficult to distinguish a high-redshift supernova against a background galaxy or from quasar activity. As a confirmation of the robustness of the pipeline used here, we note that of the 9 objects identified as outliers in all data sets of case C, 5 are miss-classifications and 1 is an extreme case of bad photometry.

In summary, the isolation forest analysis identified 81 potentially interesting objects, from which 27 (33\%) where confirmed to be non-SN events or representatives of the rare SN classes. Found anomalies correspond to 1.4\% of the original data set of $\sim$2000 objects which was identified demanding significantly less resources than a manual search would entail. Among these objects, we report for the first time the 16 star/quasar-like objects misclassified as SNe.

It is important to note that this results are not expected to be complete. For example, there are known SLSN which were not identified as outliers in our search, as well as 1 object (SN1000+0216) at very high redshift which was identified as anomalous in only 1 of the data sets for case C --- and consequently was not included in our final list. This is a natural consequence of the pre-processing analysis we chose to adopt, where the objects are mainly characterised by their light curve shape (all photometric features were normalised). In this context, differences in intrinsic brightness only marginally affect our final results. Another source of false negatives can be traced back to GP approximations, the typical example being  Gaia16aye (Fig. \ref{Gaia16aye}). From a visual inspection of its observed photometric points this objects is obviously not a SN. However, it appeared only in 3 of the 8 possible data sets in case C. A more detailed analysis of  its GP approximation (solid line in Fig. \ref{Gaia16aye}), reveals that the information input to the ML model was much smoother than one would expect. As a consequence, the algorithm struggles to separate it from other slow declining events.

Nevertheless, the above results provide clear evidence of the effectiveness of automated anomaly detection algorithms for photometric SN light curve analysis. In this work we used data from the OSC in order to provide a proof of concept. Although this is not a big data sample, it does allow us to search for independent information in the literature on which we could confirm our findings. This approach to the analysis of photometric light curves will be paramount for  future astronomical surveys like LSST, which will not be able to afford a manual research or the possibility to overlook interesting objects deviating from the bulk of the data  --- where the most interesting physics resides.

The code of this work and the data are available at\\ \href{http://snad.space/osc/}{http://snad.space/osc/}.

\section*{Acknowledgements}

M.~Pruzhinskaya and M.~Kornilov are supported by RFBR grant according to the research project 18-32-00426 for outlier analysis and LCs approximation.  K.~Malanchev is supported by RBFR grant 18-32-00553 for preparing the Open Supernova Catalog data. E.~E.~O.~Ishida acknowledges support from CNRS 2017 MOMENTUM grant and Foundation for the advancement of theoretical physics and Mathematics "BASIS". A.~Volnova acknowledges support from RSF grant 18-12-00522 for analysis of interpolated LCs. We used the equipment funded by the Lomonosov Moscow State University Program of Development. The authors acknowledge the support from the Program of Development of M.V. Lomonosov Moscow State University (Leading Scientific School "Physics of stars, relativistic objects and galaxies"). This research has made use of NASA's Astrophysics Data System Bibliographic Services and following {\sc Python} software packages: {\sc NumPy}~\citep{numpy}, {\sc Matplotlib}~\citep{matplotlib}, {\sc SciPy}~\citep{scipy}, {\sc pandas}~\citep{pandas}, and {\sc scikit-learn}~\citep{scikit-learn}.

%%%%%%%%%%%%%%%%%%%%%%%%%%%%%%%%%%%%%%%%%%%%%%%%%%

%%%%%%%%%%%%%%%%%%%% REFERENCES %%%%%%%%%%%%%%%%%%

\bibliographystyle{mnras}
\bibliography{snad_biblio} 

%%%%%%%%%%%%%%%%%%%%%%%%%%%%%%%%%%%%%%%%%%%%%%%%%%

%%%%%%%%%%%%%%%%% APPENDICES %%%%%%%%%%%%%%%%%%%%%

\appendix
\section{Table with outliers}
\label{anomaly_list}

{\scriptsize
\onecolumn
\begin{landscape}
\renewcommand*{\arraystretch}{1.7}
\begin{longtable}{lccp{1.5cm}clccp{0.3cm}p{0.3cm}p{0.3cm}p{5cm}p{2cm}}
\caption{List of outliers and their hosts}
\label{outliers_table}\\
\hline
Name &$\alpha$ &$\delta$ &Type$\rm^a$ &z$_{\rm CMB}$ &Host name &Host $\alpha$ &Host $\delta$ &Host type$\rm^b$ &Sep. ('')$\rm^c$ &Sep. (kpc)$\rm^c$ &Comments$\rm^d$ &References  \\
\hline 
\multicolumn{13}{c}{Outliers found in 8 data sets with different dimensionality reduction} \\ 
\hline
SDSS-II SN 13112$^\dagger$ &01:09:29.89   &$-$01:05:18.3   &?SN II        &       &SDSS J010929.89-010518.3     &01:09:29.89   &$-$01:05:18.3   &  &     &     &pSN II in~\cite{Sako2018}; SDSS DR15 host photoZ (KD-tree method) $0.755\pm0.234$                                                         &\cite{Sako2018}  \\
SDSS-II SN 13461 &23:03:39.48   &$+$00:12:49.9   &?SN II        &       &SDSS J230339.49+001249.7     &23:03:39.49   &$+$00:12:49.8   &  &     &     &pSN II in~\cite{Sako2018}; SDSS DR15 host photoZ (KD-tree method) $0.735\pm0.080$                                                         &\cite{Sako2018}  \\
SDSS-II SN 5314$^\dagger$  &20:21:17.85   &$+$00:41:04.3   &?SN II/?Star  &       &SDSS J202117.84+004104.2     &20:21:17.84   &$+$00:41:04.2   &  &     &     &pSN II in~\cite{Sako2018}; host classified as star by SDSS DR15                                                                           &\cite{Sako2018}  \\
SN2016fbo$^\star$        &01:01:35.54   &$+$17:06:04.3   &SN Ia         &0.030  &GALEXASC J010135.75+170604.9 &01:01:35.77   &$+$17:06:04.1   &  & 3.3   &  1.98   &LC in the Open Supernova Catalog has a bad quality                                                                                        &\cite{2018MNRAS.475..193F}  \\
SDSS-II SN 14170$^\dagger$ &01:46:33.15   &$+$00:51:05.7   &?SN II/?Star  &       &SDSS J014633.15+005105.6     &01:46:33.15   &$+$00:51:05.6   &  &     &     &pSN II in~\cite{Sako2018}; host classified as star by SDSS DR15                                                                           &\cite{Sako2018}  \\
SDSS-II SN 1706$^\dagger$  &00:02:58.11   &$-$01:01:27.8   &?SN II/QSO    &       &SDSS J000258.10-010127.8     &00:02:58.10   &$-$01:01:27.8   &  &     &     &pSN II in~\cite{Sako2018}; according to SDSS DR15 host has BOSS spectrum with $z = 1.551$, class = QSO broadline                          &\cite{Sako2018}  \\
LSQ13dpa$^\dagger$         &11:01:12.91   &$-$05:50:52.6   &SN II         &0.023  &LCSB S1492O                  &11:01:12.46   &$-$05:50:46.0   &S &  9.4   &  4.37   &Spectroscopically confirmed as SN~II using a near-infrared spectrum (range 800-2500 nm)                                                   &\cite{2013ATel.5678....1H}  \\
SDSS-II SN 17756$^\dagger$ &01:08:10.42   &$-$00:16:36.9   &?SN II/QSO    &       &SDSS J010810.43-001636.9     &01:08:10.43   &$-$00:16:37.0   &  &     &     &pSN II in~\cite{Sako2018}; host classified as star by SDSS DR15, however, it has a BOSS spectrum with $z = 1.997$, class = QSO broadline  &\cite{Sako2018}  \\
SDSS-II SN 15565 &01:00:27.12   &$+$00:35:23.6   &?SN II/?Star  &       &SDSS J010027.10+003523.5     &01:00:27.10   &$+$00:35:23.5   &  &     &     &pSN II in~\cite{Sako2018}; host classified as star by SDSS DR15                                                                           &\cite{Sako2018}  \\
\hline 
\multicolumn{13}{c}{Outliers found in 7 data sets with different dimensionality reduction} \\ 
\hline
SN2005ho$^\star$         &00:59:24.10 &$+$00:00:09.3 &SN Ia       &0.062 &PGC 1154577              &00:59:24.10 &$+$00:00:09.4 &Sm/Im & 0.1 & 0.12 &In JLA cosmological sample~\citep{Betoule2014}, not in Pantheon~\citep{2018ApJ...859..101S}                                     &\cite{Betoule2014}  \\
SN2016ija        &04:12:07.62 &$-$32:51:10.9 &SN II       &0.003 &NGC 1532                 &04:12:04.33 &$-$32:52:27.2 &Sbc   & 86.8 & 5.39 &Highly obscured SN II ($E(B-V)_{\rm host} = 1.95\pm0.15$~mag)                                                                                                          &\cite{2018ApJ...853...62T}  \\
SDSS-II SN 2050$^\dagger$  &20:54:41.52 &$-$00:13:33.2 &Unknown     &      &SDSS J205441.53-001333.1 &20:54:41.53 &$-$00:13:33.2 &      & & &Unknown object in~\cite{Sako2018} and SN II in the Open Supernova Catalog with reference to~\cite{Sako2018}                     &\cite{Sako2018}  \\
SDSS-II SN 17339$^\dagger$ &02:31:22.22 &$+$00:23:09.6 &?SN II/QSO  &      &SDSS J023122.22+002309.5 &02:31:22.22 &$+$00:23:09.5 &      & & &pSN II in~\cite{Sako2018}; host classified as star by SDSS DR15, however, it has a BOSS spectrum with $z = 1.132$, class = QSO  &\cite{Sako2018}  \\
\hline 
\multicolumn{13}{c}{Outliers found in 6 data sets with different dimensionality reduction} \\ 
\hline
SN2007jm$^{\star,\dagger}$         &21:55:38.59 &$-$00:10:36.3 &SN IIn       &0.090  &SDSS J215538.80-001034.1 &21:55:38.80 &$-$00:10:34.2   &      & 3.8 & 6.36 &According to SDSS DR15 host has BOSS spectrum with  $z = 0.091$, class = galaxy starforming                       &\cite{2007CBET.1079....1B,Sako2018}  \\
SNLS-06D3gx      &14:17:03.23 &$+$52:56:10.5 &SN Ia        &0.761  &SDSS J141703.21+525616.1 &14:17:03.21 &$+$52:56:16.1   &      & 5.6 & 41.31 &In JLA cosmological sample~\citep{Betoule2014}, not in Pantheon~\citep{2018ApJ...859..101S}                       &\cite{Betoule2014}  \\
SDSS-II SN 13725 &21:06:55.01 &$+$00:53:44.9 &?SN II/?Star &      &SDSS J210655.01+005344.8 &21:06:55.01 &$+$00:53:44.9   &      & & &pSN II in~\cite{Sako2018}; host classified as star by SDSS DR15                       &\cite{Sako2018}  \\
SN2016ixf        &10:39:44.56 &$+$15:02:04.8 &SN Ia        &0.067  &SDSS J103944.53+150204.7 &10:39:44.53 &$+$15:02:04.8   &      & 0.4 & 0.56 &                     &\cite{2016ATel.9889....1C,2018MNRAS.475..193F}  \\
SN2006ob         &01:51:48.14 &$+$00:15:47.9 &SN Ia        &0.059  &UGC 1333                 &01:51:48.51 &$+$00:15:49.8   &Sa    & 5.9 & 6.69 &In JLA~\citep{Betoule2014} and Pantheon~\citep{2018ApJ...859..101S} cosmological samples                       &\cite{Betoule2014}  \\
SDSS-II SN 17509 &00:19:18.93 &$+$01:08:14.3 &?SN II/QSO   &       &SDSS J001918.93+010814.2 &00:19:18.93 &$+$01:08:14.2   &      & & &pSN II in~\cite{Sako2018}; host classified as star by SDSS DR15, however, it has a BOSS spectrum with $z = 2.031$, class = QSO                       &\cite{Sako2018}  \\
\hline 
\multicolumn{13}{c}{Outliers found in 5 data sets with different dimensionality reduction} \\ 
\hline
SDSS-II SN 4330  &01:44:35.82 &$-$00:10:57.4   &?SN II       &       &SDSS J014435.82-001057.3 &01:44:35.82 &$-$00:10:57.4     &     &  & &pSN II in~\cite{Sako2018}; SDSS DR15 host photoZ (KD-tree method) $0.732\pm0.076$                       &\cite{Sako2018}  \\
SN2005ll         &22:28:06.87 &$-$01:07:41.4   &SN Ia        &0.241  &SDSS J222806.92-010742.1 &22:28:06.92 &$-$01:07:42.1     &    & 1.0 & 3.90 &According to SDSS DR15 host has BOSS spectrum with  $z = 0.242$, class = galaxy starforming                       &\cite{Sako2018}  \\
SDSS-II SN 13741 &20:48:00.40 &$-$01:02:49.5   &?SN II/?Star &       &SDSS J204800.39-010249.4 &20:48:00.39 &$-$01:02:49.5     &     &  & &pSN II in~\cite{Sako2018}; host classified as star by SDSS DR15                       &\cite{Sako2018}  \\
SDSS-II SN 17292$^\dagger$ &23:31:23.77 &$+$00:37:45.6   &?SN II       &       &SDSS J233123.77+003745.4 &23:31:23.77 &$+$00:37:45.5     &     &  & &pSN II in~\cite{Sako2018}; SDSS DR15 host photoZ (KD-tree method) $0.595\pm0.163$                       &\cite{Sako2018}  \\
SN2006kg         &01:04:16.98 &$+$00:46:08.9   &AGN          &0.230  &SDSS J010416.98+004608.7 &01:04:16.98 &$+$00:46:08.8     &     &  & &Basing on NTT spectrum classified as AGN by \cite{2011AA...526A..28O}; according to SDSS DR15 host has BOSS spectrum with $z = 0.231$, class = galaxy starburst                       &\cite{Sako2018}  \\
\hline 
\multicolumn{13}{c}{Outliers found in 4 data sets with different dimensionality reduction} \\ 
\hline
SDSS-II SN 4652  &02:29:49.69 &$-$00:40:11.4     &?SN II/QSO          &       &SDSS J022949.69-004011.3 &02:29:49.69  &$-$00:40:11.4 &   & & &pSN II in~\cite{Sako2018}; according to SDSS DR15 host has BOSS spectrum with  $z = 0.673$, class = QSO                       &\cite{Sako2018}  \\
SN2002bj         &05:11:46.41 &$-$15:08:10.8     &SN Ia pec/SN Ib pec &0.012  &NGC 1821                 &05:11:46.11  &$-$15:08:04.9 &Im & 7.3 & 1.80 &Bright, fast-evolving supernova with low-mass ejecta, helium and carbon lines in spectra                       &\cite{2010Sci...327...58P}  \\
SDSS-II SN 13589$^\dagger$ &21:48:02.39 &$-$00:07:07.5     &?SN II              &       &SDSS J214802.31-000710.0 &21:48:02.31  &$-$00:07:10.1 &   & & &pSN II in~\cite{Sako2018}; SDSS DR15 host photoZ (KD-tree method) $0.199\pm0.032$                       &\cite{Sako2018}  \\
SDSS-II SN 13291 &20:04:11.38 &$-$00:32:01.1     &?SN II              &       &SDSS J200411.38-003200.9 &20:04:11.38  &$-$00:32:01.0 &   & & &pSN II in~\cite{Sako2018}; SDSS DR15 host photoZ (KD-tree method) $0.425\pm0.119$                       &\cite{Sako2018}  \\
SN2213-1745      &22:13:39.97 &$-$17:45:24.5     &SLSN-R              &2.046  &                         &             &              &   & & &                     &\cite{2012Natur.491..228C}  \\
SN2017mf         &14:16:31.00 &$+$39:35:12.0     &SN Ia               &0.026  &NGC 5541                 &14:16:31.80  &$+$39:35:20.7 &Sb & 12.7 & 6.64 &                     &\cite{2018MNRAS.475..193F}  \\
SN2017yh         &17:52:06.25 &$+$21:33:58.3     &SN Ia               &0.020  &IC 1269                  &17:52:05.86  &$+$21:34:09.0 &Sbc& 12.0 & 4.86 &                     &\cite{2018MNRAS.475..193F}  \\
SN2013cv         &16:22:43.19 &$+$18:57:35.0     &SN Ia pec           &0.036  &SDSS J162243.02+185733.8 &16:22:43.02  &$+$18:57:33.8 &   & 2.7 & 1.93 &Large peak optical and UV luminosity, absence of iron absorption lines in the early spectra                       &\cite{0004-637X-823-2-147}  \\
SN2006pt$^\star$         &02:27:16.17 &$-$00:23:36.5     &SN Ia               &0.298  &SDSS J022716.08-002335.6 &02:27:16.08  &$-$00:23:35.6 &   & 1.6 & 7.21 &According to SDSS DR15 host has BOSS spectrum with  $z = 0.299$, class = galaxy starforming                       &\cite{Sako2018}  \\
SDSS-II SN 2661  &23:32:49.80 &$+$00:05:50.0     &SN II               &0.191  &SDSS J233249.88+000549.2 &23:32:49.88  &$+$00:05:49.2 &   & 1.4 & 4.59 &                     &\cite{Sako2018}  \\
SDSS-II SN 20266 &00:31:13.40 &$-$00:07:08.7     &?SN II              &       &                         &             &              &   & & &pSN II in~\cite{Sako2018}                       &\cite{Sako2018}  \\
\hline 
\multicolumn{13}{c}{Outliers found in 3 data sets with different dimensionality reduction} \\ 
\hline
SDSS-II SN 12868 &21:29:40.40 &$-$00:01:38.8     &?SN II              &       &SDSS J212940.40-000138.9     &21:29:40.40  &$-$00:01:39.0 &   &  &  &pSN II in~\cite{Sako2018}; SDSS DR15 host photoZ (KD-tree method) $0.688\pm0.167$                       &\cite{Sako2018}  \\
SDSS-II SN 19699 &02:02:11.76 &$+$00:13:46.3     &?SN II/?Star        &       &SDSS J020211.76+001346.2     &02:02:11.76  &$+$00:13:46.2 &   &  &  &pSN II in~\cite{Sako2018}; host classified as star by SDSS DR15                       &\cite{Sako2018}  \\
SDSS-II SN 16302 &22:07:04.15 &$+$00:11:00.6     &?SN Ia              &       &                             &22:07:04.11  &$+$00:10:58.9 &   &  &  &pSN II in~\cite{Sako2018}; host photoZ $0.185\pm0.015$~\citep{2012ApJ...755...61S}                       &\cite{Sako2018}  \\
SDSS-II SN 15745 &02:48:49.91 &$-$00:06:27.1     &?SN Ia              &       &SDSS J024849.89-000626.6     &02:48:49.89  &$-$00:06:26.7 &   &  &  &pSN Ia in~\cite{Sako2018}; SDSS DR15 host photoZ (KD-tree method) $0.657\pm0.074$                       &\cite{Sako2018}  \\
Gaia16aye$^\star$        &19:40:01.10 &$+$30:07:53.4     &ULENS, CV           &       &MW                           &             &              &   &  &  &Binary microlensing event                        &\cite{2016ATel.9376....1B}  \\
PS1-1000007      &02:23:30.71 &$-$04:38:10.8     &SN Ia               &0.137  &SDSS J022330.91-043810.6     &02:23:30.91  &$-$04:38:10.7 &   & 3.0 & 7.25 &                     &\cite{2014ApJ...795...44R}  \\
SN2006ne         &01:13:37.84 &$+$00:25:26.0     &SN Ia               &0.046  &SDSS J011337.58+002525.5     &01:13:37.58  &$+$00:25:25.5 &S  & 3.9 & 3.55 &According to SDSS DR15 host has BOSS spectrum with  $z = 0.047$, class = galaxy starforming                       &\cite{Sako2018}  \\
SDSS-II SN 19504 &00:54:39.68 &$-$00:25:23.9     &SN II               &0.214  &SDSS J005439.88-002526.3     &00:54:39.88  &$-$00:25:26.3 &   & 3.8 & 13.36 &                     &\cite{Sako2018}  \\
SDSS-II SN 18266 &03:15:24.35 &$-$00:50:54.0     &?SN II/?Star        &       &SDSS J031524.34-005053.9     &03:15:24.34  &$-$00:50:54.0 &   &  &  &pSN II in~\cite{Sako2018}; host classified as star by SDSS DR15                       &\cite{Sako2018}  \\
PS15cfn          &21:59:21.97 &$-$21:07:10.7     &91T-like            &0.109  &GALEXASC J215922.04-210713.3 &21:59:22.00  &$-$21:07:13.5 &   & 2.8 & 5.63 &                     &\cite{2018MNRAS.475..193F}  \\
SDSS-II SN 15048 &02:54:57.10 &$-$00:00:18.6     &?SN Ia              &       &                             &             &              &   &  &  &pSN Ia in~\cite{Sako2018}                       &\cite{Sako2018}  \\
\hline 
\multicolumn{13}{c}{Outliers found in 2 data sets with different dimensionality reduction} \\ 
\hline
SN2006ej          &00:38:59.77  &$-$09:00:56.6  &SN Ia           &0.020    &IC 1563                        &00:39:00.24   &$-$09:00:52.4 &S0       & 8.1 & 3.30 &In JLA~\citep{Betoule2014} and Pantheon~\citep{2018ApJ...859..101S} cosmological samples                       &\cite{Betoule2014}  \\
SDSS-II SN 18391  &02:22:42.43  &$+$00:25:05.0  &?Unknown/?Star  &         &SDSS J022242.43+002504.8       &02:22:42.43   &$+$00:25:04.9 &         &  & &Unknown object in~\cite{Sako2018}; host classified as star by SDSS DR15                       &\cite{Sako2018}  \\
SN1996ai          &13:10:58.13  &$+$37:03:35.4  &SN Ia           &0.004    &NGC 5005                       &13:10:56.31   &$+$37:03:32.2 &Sbc      & 22.0 & 1.82 &Highly reddened SN Ia ($ E(B-V)_{\rm host}=1.69\pm0.10$ mag)                       &\cite{2007ApJ...659..122J,2008ApJ...675..626W}  \\
SN2016ayg         &07:30:17.40  &$+$25:01:56.0  &SN Ia           &0.043    &SDSS J073017.25+250153.5       &07:30:17.26   &$+$25:01:53.5 &         & 3.1 & 2.66 &                     &\cite{2018MNRAS.475..193F}  \\
SDSS-II SN 4226   &23:40:41.66  &$-$00:54:21.2  &?SN II/?Star    &         &SDSS J234041.66-005421.3       &23:40:41.66   &$-$00:54:21.3 &         &  & &pSN II in~\cite{Sako2018}; host classified as star by SDSS DR15                       &\cite{Sako2018}  \\
SN2005jw          &20:40:19.25  &$-$00:00:25.8  &SN Ia           &0.380    &SDSS J204019.14-000022.8       &20:40:19.14   &$-$00:00:22.9 &Sbc/Sc   & 3.3 & 17.37 &In JLA~\citep{Betoule2014} and Pantheon~\citep{2018ApJ...859..101S} cosmological samples                       &\cite{2008AJ....135.1766Z,Betoule2014}  \\
SN2016bln$^\star$         &13:34:45.49  &$+$13:51:14.3  &91T-like        &0.024    &NGC 5221                       &13:34:55.91   &$+$13:49:57.1 &Sb       & 170.3 & 82.41 &Peculiar  rise  time, non-evolving blue colour, unusual strong $\rm C~II$ absorption                        &\cite{2018MNRAS.475..193F}  \\
SN2016bmc         &19:10:37.33  &$+$37:39:17.0  &SN Ia           &0.028    &UGC 11409                      &19:10:37.51   &$+$37:39:18.9 &S        & 2.9 & 1.61 &                     &\cite{2018MNRAS.475..193F}  \\
SDSS-II SN 2093   &22:36:36.28  &$-$00:12:47.0  &?SN II          &         &                               &              &              &         &  & &pSN II in~\cite{Sako2018}                       &\cite{Sako2018}  \\
SDSS-II SN 17317  &01:29:59.18  &$-$00:38:05.4  &?SN II          &         &SDSS J012959.31-003800.3       &01:29:59.31   &$-$00:38:00.3 &Sbc      &  & &zSN II in~\cite{Sako2018}; according to SDSS DR15 host has BOSS spectrum with  $z = 0.118$, class = starforming galaxy                       &\cite{Sako2018}  \\
SNLS-03D1cm       &02 24 55.27  &$-$04 23 03.4  &91T-like        &0.870    &[HSP2005] J022455.28-042303.68 &02:24:55.28   &$-$04:23:03.7 &         & 0.3 & 2.59 &Peculiar Type Ia SN with the stretch-related parameter $X_1=4.54$                     &\cite{2010AA...523A...7G,2011AA...534A..43B}  \\
SDSS-II SN 17789  &01:29:16.13  &$+$00:42:37.9  &SLSN            &         &                               &              &              &         &  & &According to table~2 of~\cite{Sako2018} SN has 4 spectra                       &\cite{Sako2018}  \\
SN2015bn          &11:33:41.57  &$+$00:43:32.2  &SLSN-I          &0.114    &SDSS J113341.53+004333.2       &11:33:41.53   &$+$00:43:33.3 &         & 1.3 & 2.59 &Hydrogen-poor superluminous supernova                       &\cite{2015ATel.7102....1L,2016ApJ...828L..18N}  \\
SDSS-II SN 2809$^\dagger$   &03:33:27.41  &$+$00:16:10.7  &?SN II/?Star    &        &SDSS J033327.41+001610.7       &03:33:27.41   &$+$00:16:10.7 &         &  & &pSN II in~\cite{Sako2018}; host classified as star by SDSS DR15                       &\cite{Sako2018}  \\
\hline 
\multicolumn{13}{c}{Outliers found in a data set of 364 photometric characteristics ($121\times3$ normalized fluxes and the LC flux maximum)} \\ 
\hline
SDSS-II SN 18228 &22:45:49.70 &$-$00:15:54.9 &?SN II        &       &                          &            &              &    &    &   &pSN II in~\cite{Sako2018}                       &\cite{Sako2018}  \\
SDSS-II SN 18733 &01:22:42.61 &$-$00:02:48.4 &?SN II        &       &                          &            &              &    &    &   &pSN II in~\cite{Sako2018}                       &\cite{Sako2018}  \\
SDSS-II SN 19047 &21:43:18.71 &$+$00:32:21.7 &?SN II        &       &SDSS J214318.74+003219.8  &21:43:18.74 &$+$00:32:19.9 &    &    &   &zSN II in~\cite{Sako2018}; according to SDSS DR15 host has BOSS spectrum with  $z = 0.412$, class =  galaxy starburst                       &\cite{Sako2018}  \\
SDSS-II SN 19395 &02:44:37.90 &$+$00:46:32.1 &?SN II/QSO    &       &SDSS J024437.89+004631.9  &02:44:37.89 &$+$00:46:32.0 &    &    &   &pSN II in~\cite{Sako2018}; according to SDSS DR15 host has BOSS spectrum with  $z = 1.318$, class = QSO                       &\cite{Sako2018}  \\
SDSS-II SN 6992  &22:10:25.09 &$+$00:00:02.7 &?SN II/?Star  &       &SDSS J221025.08+000002.5  &22:10:25.08 &$+$00:00:02.5 &    &    &   &pSN II in~\cite{Sako2018}; host classified as star by SDSS DR15                       &\cite{Sako2018}  \\
SN2005mp         &01:04:45.68 &$+$00:03:20.2 &SN Ia         &0.272  &                          &            &              &    &    &   &Host identified by \citealt{Sako2018} (SDSS J010445.51+000320.8) has BOSS spectrum with $z = 0.952$ that is different from SN redshift                       &\cite{2013MNRAS.433.2240G,Sako2018}  \\
SN2013ab$^\dagger$         &14:32:44.49 &$+$09:53:12.3 &SN IIP        &0.006  &NGC 5669                  &14:32:43.80 &$+$09:53:28.8 &Scd &  19.4  & 2.40  &The light curve and spectra suggest that the supernova is a normal Type IIP event, although with a steeper decline during the plateau relative to other archetypal SNe of similar brightness                       &\cite{2015MNRAS.450.2373B}  \\
SN1999gi         &10:18:16.66 &$+$41:26:28.2 &SN IIP        &0.002  &NGC 3184                  &10:18:16.99 &$+$41:25:27.8 &Sc  &  60.5  & 2.51  &                        &\cite{2002AJ....124.2490L}  \\
\hline 
\multicolumn{13}{c}{Outliers found in a data set of 10 Gaussian process parameters (9 fitted parameters of the kernel and the log-likelihood of the fit)} \\ 
\hline
PTF10aagc    &09:39:56.93 &$+$21:43:16.9 &SLSN-I         &0.207  &SDSS J093956.91+214317.1     &09:39:56.91  &$+$21:43:17.1  &     & 0.3 & 1.16 &SLSN-I with hydrogen in late spectra; host morphological structure suggests a possible ongoing merger                       &\cite{2016ApJ...830...13P,2018ApJ...860..100D}  \\
SN2006T      &09:54:30.21 &$-$25:42:29.3 &SN IIb         &0.008  &NGC 3054                     &09:54:28.61  &$-$25:42:12.4  &Sbc  & 27.4 & 4.51 &                     &\cite{2006IAUC.8666....2M,2018AA...609A.134S}  \\
SN2010bb     &10:44:38.23 &$+$57:48:40.0 &SN Ia          &0.118  &SDSS J104438.19+574839.8     &10:44:38.19  &$+$57:48:39.8  &     & 0.4 & 0.80 &                     &\cite{2014ApJ...795...44R}  \\
SN2013ej     &01:36:48.16 &$+$15:45:31.0 &SN IIP/SN IIL  &0.002  &NGC 628                      &01:36:41.77  &$+$15:47:00.5  &Sc   & 128.5 & 5.32 &LC is intermediate between those of Type IIP and IIL SNe                      &\cite{2015ApJ...806..160B,2015ApJ...807...59H,2017ApJ...834..118M}  \\
SNLS-04D2gb  &10:02:22.67 &$+$01:53:39.0 &SN Ia          &0.452  &SDSS J100222.66+015339.2     &10:02:22.66  &$+$01:53:39.2  &     & 0.2 & 1.44 &In JLA~\citep{Betoule2014} and Pantheon~\citep{2018ApJ...859..101S} cosmological samples; host classified as star by SDSS DR15                       &\cite{Betoule2014}  \\
SNLS-05D2mp  &09:59:08.63 &$+$02:12:14.4 &SN Ia          &0.355  &                             &             &               &     & & &In JLA~\citep{Betoule2014} and Pantheon~\citep{2018ApJ...859..101S} cosmological samples                       &\cite{Betoule2014}  \\
SNLS-06D2ag  &10:01:43.36 &$+$01:51:37.1 &SN Ia          &0.310  &SDSS J100143.26+015135.4     &10:01:43.26  &$+$01:51:35.4  &     & 2.3 & 10.33 &                     &\cite{2010AA...523A...7G,2013MNRAS.433.2240G}  \\
SNLS-06D3cn  &14:19:25.85 &$+$52:38:27.5 &SN Ia          &0.232  &SDSS J141925.79+523825.9     &14:19:25.79  &$+$52:38:25.9  &     & 1.7 & 6.25 &                     &\cite{2010AA...523A...7G}  \\
SN1999cc     &16:02:42.03 &$+$37:21:34.4 &SN Ia          &0.032  &NGC 6038                     &16:02:40.55  &$+$37:21:34.2  &Sbc  & 17.6 & 11.28 &In JLA~\citep{Betoule2014} and Pantheon~\citep{2018ApJ...859..101S} cosmological samples                       &\cite{Betoule2014}  \\
SN2002aw     &16:37:29.06 &$+$40:52:50.3 &SN Ia          &0.026  &2MFGC 13321                  &16:37:29.22  &$+$40:52:48.2  &Sb   & 2.8 & 1.45 &                     &\cite{2013MNRAS.433.2240G}  \\
SN2002eb     &22:19:05.24 &$+$24:35:39.8 &SN Ia          &0.026  &PGC 68560                    &22:19:06.29  &$+$24:35:53.4  &Sb   & 19.7 & 10.33 &                     &\cite{2013MNRAS.433.2240G}  \\
SN2004dt     &02:02:12.77 &$-$00:05:51.5 &SN Ia          &0.019  &NGC 799                      &02:02:12.30  &$-$00:06:02.6  &Sa   & 13.1 & 5.07 &Spectral subtype: high velocity (HV, \citealt{Wang2009}), broad line (BL, \citealt{Branch2006})                       &\cite{2013ApJ...773...53F}  \\
SN2009bw     &03:56:06.92 &$+$72:55:40.9 &SN IIP         &0.004  &UGC 2890                     &03:56:04.44  &$+$72:55:18.5  &Sdm  & 24.9 & 2.06 &Luminosity drop from the photospheric to the nebular phase is one of the fastest ever observed, $\sim$2.2 mag in $\sim$13 days                       &\cite{2012MNRAS.422.1122I,2014AJ....148..107R}  \\
  \hline
\multicolumn{13}{l}{$\rm ^a$ Type of the source. A prefix "?" means that the source is not confirmed spectroscopically.} \\
\multicolumn{13}{l}{$\rm ^b$ Simbad host galaxy morphological type.} \\
\multicolumn{13}{l}{$\rm ^c$ Separation of the source from the center of its host galaxy.} \\
\multicolumn{13}{l}{$\rm ^d$  If classification is made by~\citealt{Sako2018}: a prefix "p" (pSN) indicates a purely photometric type, a prefix "z" (zSN) indicates that a redshift is measured from its candidate host galaxy and the classification uses that redshift as a prior.} \\
\multicolumn{13}{l}{$^\star$ The object is also found in a data set of 10 Gaussian process parameters (9 fitted parameters of the kernel and  the log-likelihood of the fit).} \\
\multicolumn{13}{l}{$^\dagger$ The object is also found in a data set of 364 photometric characteristics ($121\times3$ normalized fluxes and the LC flux maximum).} \\
\end{longtable}
% \end{landscape}
% }

%%%%%%%%%%%%%%%%%%%%%%%%%%%%%%%%%%%%%%%%%%%%%%%%%%

% Don't change these lines
\bsp	% typesetting comment
\label{lastpage}
\end{landscape}
}

\end{document}